\documentclass[a4paper,11pt]{article}
\usepackage{etex}      

\pdfoutput=1 

\usepackage{jheppub} 
\usepackage{subfig}						
\usepackage[table]{xcolor}				
\usepackage{colortbl}
\usepackage{booktabs}					
\usepackage{multirow,bigdelim}			
\usepackage{longtable}					
\usepackage{arydshln}					
\usepackage{units}        
\usepackage{amsmath,amssymb,amsthm,amscd,graphicx}
\usepackage{psfrag}
\input epsf.sty

\usepackage{color}

\theoremstyle{definition}

\newcommand{\CM}{{\cal M}}
\newcommand{\CN}{{\cal N}}

\newcommand{\CW}{{\cal W}}

\def\IN{{\mathbb N}}
\def\IZ{{\mathbb Z}}
\def\IR{{\mathbb R}}
\def\IC{{\mathbb C}}
\def\IP{{\mathbb P}}

\def\IF{{\mathbb F}}
\def\IQ{{\mathbb Q}}

\newcommand{\re}{{\rm e}}
\newcommand{\ri}{{\mathsf{i}\,}}

\newcommand{\Li}{\mathop{\rm Li}\nolimits}

\newcommand{\cC}{{\mathcal{C}}}

\newcommand{\dB}{\mathbf{B}}
\newcommand{\dC}{\mathbf{C}}
\newcommand{\dt}{\mathbf{t}}
\newcommand{\dn}{\mathbf{n}}

\newcommand{\mfr}{\mathfrak}
\newcommand{\bb}{\boldsymbol}

\newcommand{\be}{\begin{equation}}
\newcommand{\ee}{\end{equation}}
\newcommand{\ba}{\begin{aligned}}
\newcommand{\ea}{\end{aligned}}
\newcommand{\ben}{\begin{eqnarray}\displaystyle}
\newcommand{\een}{\end{eqnarray}}

\newdimen\tableauside\tableauside=1.0ex
\newdimen\tableaurule\tableaurule=0.4pt
\newdimen\tableaustep
\def\phantomhrule#1{\hbox{\vbox to0pt{\hrule height\tableaurule width#1\vss}}}
\def\phantomvrule#1{\vbox{\hbox to0pt{\vrule width\tableaurule height#1\hss}}}
\def\sqr{\vbox{%
  \phantomhrule\tableaustep
  \hbox{\phantomvrule\tableaustep\kern\tableaustep\phantomvrule\tableaustep}%
  \hbox{\vbox{\phantomhrule\tableauside}\kern-\tableaurule}}}
\def\squares#1{\hbox{\count0=#1\noindent\loop\sqr
  \advance\count0 by-1 \ifnum\count0>0\repeat}}
\def\tableau#1{\vcenter{\offinterlineskip
  \tableaustep=\tableauside\advance\tableaustep by-\tableaurule
  \kern\normallineskip\hbox
    {\kern\normallineskip\vbox
      {\gettableau#1 0 }%
     \kern\normallineskip\kern\tableaurule}%
  \kern\normallineskip\kern\tableaurule}}
\def\gettableau#1{\ifnum#1=0\let\next=\null\else
\squares{#1}\let\next=\gettableau\fi\next}

\graphicspath{{figs/}}
\def\fq{\mathfrak{q}}
\def\cl{\mathcal}
\def\({\left(}
\def\){\right)}

\def\mbb{\mathbb}
\def\mbf{\mathbf}
\def\mcl{\mathcal}

\def\mfr{\mathfrak}

\newcommand{\nn}{\nonumber \\}

\title{\boldmath BPS relations from  spectral problems and  blowup equations}
\author{Alba Grassi$^{a}$ and Jie Gu$^{b}$ }
\affiliation{
$^a$International Center for Theoretical Physics,\\
 ICTP, Strada Costiera 11, Trieste 34151,Italy \\
 \\
 $^b$Laboratoire de Physique Th\'eorique de l'\'{E}cole Normale Sup\'erieure\\
 CNRS, PSL Research University, Sorbonne Universit\'{e}s, UPMC, 75005 Paris, France \\}

\emailAdd{agrassi@ictp.it, gu@lpt.ens.fr}

 \abstract{Recently an exact duality  between topological string and the spectral theory of operators constructed from mirror curves to toric Calabi-Yau threefolds has been proposed. At the same time an exact quantization condition for the cluster integrable systems associated to  these geometries has been conjectured. The consistency between the two approaches leads to an infinite set of constraints for the refined BPS invariants of  the toric Calabi-Yau threefolds. We prove these constraints for the $Y^{N,m}$ geometries using the $K$-theoretic blowup equations for $SU(N)$ SYM with generic Chern-Simons invariant $m$.}


\begin{document}
\maketitle
\flushbottom

\section{Introduction}

	It is now well-known that a class of four dimensional $\CN=2$ supersymmetric field theories have an underlying algebraically integrable system, i.e. a Hitchin system \cite{Donagi:1995cf}. 
	If we subject the four dimensional theory to the $\Omega$-background with only one deformation parameter $\epsilon_1$ turned on, the deformed supersymmetric field theory is described by the quantization of the classical integrable system where $\epsilon_1$ plays the role of the Planck constant. The supersymmetric vacua correspond to the eigenstates of the quantum many-body system, which could be solved by the following  Bethe-ansatz type equation
	\begin{equation}\label{eq:BA}
		\frac{\partial {\CW}(\vec{a},\hbar)}{\partial a_i} = 2\pi(n_i +1/2 )  \ , \quad n_i \in \IZ_{\geqslant 0} \ ,
	\end{equation}
	where $a_i$ are action variables in the integrable system, or the Coulomb moduli if interpreted from the $\CN=2$ theory side, and $\CW(\vec{a},\hbar)$ is the analogue of the Yang-Yang function. It was pointed out in \cite{ns} that $\CW(\vec{a},\hbar)$ is given by the Nekrasov-Shatashvili (NS) limit of the Nekrasov partition function. 	
	
	It is beneficial to consider the $4d$ $\CN=2$ theory as a $5d$ $\CN=1$ theory on $S^1$ in the limit where the radius of $S^1$ shrinks to zero. The partition function of the $5d$ theory is given by the $K$-theoretic Nekrasov partition function while  the $5d$ theory	
	corresponds to the relativistic lift of the integrable system that underlies the $4d$ theory \cite{Nekrasov:1996cz}. For instance, $5d$ $\CN=1$ pure SYM is associated to the relativistic periodic Toda lattice \cite{Ruijsenaars:1990toda}. Generalizing the $4d$ story, one expects that the eigenvalues of the quantization of the relativistic integrable system are solved from the same  Bethe-ansatz type equation \eqref{eq:BA} where $\CW(\vec{a},\hbar)$ is given by the NS free energy $F_{\rm NS}(\vec{a},\hbar)$ of the 5d $\CN=1$ theory  \cite{mirmor,acdkv}. 

	 However this proposal, which is based on a perturbative analysis, cannot be the full answer and some non--perturbative effects have to be added. Indeed the  5d NS free energy $F_{\rm NS}(\vec{a},\hbar)$ is singular when $\hbar \in 2\pi\IQ$, while 
	 the energy levels of the corresponding  operators are well defined 
	 positive numbers for all positive values of $\hbar$ \cite{km}. The reason for this inconsistency is revealed to be the missing of important non--perturbative contributions to the quantization condition \eqref{eq:BA}, in this case complex instantons \cite{km}. This phenomenon is similar to the one observed in the study of the quartic oscillator  \cite{bpv2,bpv1,voross} where it was pointed out that only when the complex instanton contributions are included does one obtain the correct spectrum of the system.	 
	In the case of quantizing the integrable system underlying a $5d$ $\mcl N=1$ theory, a necessary requirement is that the  
	 added non--perturbative contributions have to cancel the poles of $F_{\rm NS}(\vec{a},\hbar)$ at $\hbar\in 2\pi\IQ$ \cite{km} similar to the HMO mechanism in the ABJM free energy \cite{hmo2}. One the other hand, as pointed out in \cite{hw}, the cancellation condition is not always sufficient, and a precise conjecture including additional  non--perturbative contributions was proposed in \cite{ghm,ghmabjm}. A notable feature of the conjecture is that the moduli $\vec{a}$ must be twisted by the so-called $\mathbf{B}$-field in the instanton contributions to the free energy $F_{\rm NS}(\vec{a},\hbar)$. Furthermore it was later shown \cite{wzh} that the quantization condition \eqref{eq:BA} with the full set of non--perturbative corrections can be reorganized in a form which is invariant under the S-transformation $\hbar \mapsto 1/\hbar$. This proposal is consistent with the conjecture \cite{Kharchev:2001rs} that the spectral problem of the quantum relativistic Toda lattice can be reduced to the representations of the ``modular dual'' of the quantum group $U_q(\mfr g)$ \cite{faddeev,Faddeev:2014qya}. In \cite{hm, fhma} this S-invariant quantization condition was generalized first to relativistic periodic Toda lattice and then to a large category of quantum ``relativistic'' integrable systems, the cluster integrable system (CIS) \cite{Goncharov2011} constructed from an arbitrary 2d convex Newton polygon.

	A related and very fruitful line of research is the study of  the quantization of  the mirror curve to a toric Calabi-Yau threefold $X_N$ associated to a Newton polygon $N$
	\cite{adkmv}. Recently, this approach has been used to propose a non--perturbative definition of topological string theory on 
	toric background \cite{ghm,cgm2} in terms of an ideal Fermi gas. According to this construction
	one can extract $g$ trace class operators, $g$ being the genus of the mirror curve, and define for them a generalized Fredholm determinant $\Xi_{X_N}$ whose explicit expression is given in term of enumerative invariants of  $X_N$ \cite{ghm,cgm2}. The exact spectrum of these operators is computed by the vanishing locus of $\Xi_{X_N}$.
	Even though a rigorous mathematical proof is still lacking, this conjecture and its generalized form have  passed numerous tests and found broad applications \cite{kama,mz,kmz,gkmr,oz,wzh,hel,
	hm,fhma,bgt,amirnew,ag,hkt2,mz2,huang1606,sug,cgum}.

	It is an intriguing observation that the quantum mirror curve to $X_N$ coincides with the quantum Baxter equation of the CIS based on $N$ \cite{fhma,huang1606,cgum}. One of the consequences of this observation is that the energy levels of the CIS must lie on the zero loci of the generalized Fredholm determinant \cite{ghm, cgm2}. This imposes a consistency condition between the two approaches demanding that there must be a suitable B-field such that  an identity which generates infinitely many constraints for the BPS invariants of $X_N$ is satisfied \cite{huang1606}. We will name this identity the {\it{compatibility formula}}. It is further conjectured that  
	 there  are at least $g$ inequivalent sets of $\mathbf{B}$-fields for which the compatibility formula holds  \cite{huang1606}, which we will refer as the {\it sufficiency conjecture}.

	In this note we give a proof of the sufficiency conjecture and the compatibility formula when $X_N$ is the resolution of the cone over the  $Y^{N,m}$ singularity (see \cite{Brini:2008rh} for a detailed description of these geometries). We will show that the compatibility formula is a natural consequence of the $K$-theoretic blowup equation \cite{Gottsche:2006bm,Nakajima:2003pg,naga,ny,Nakajima:2011}. Furthermore, we give a compact formula for all the inequivalent $\mathbf{B}$-fields that satisfy the compatibility formula.
	
	The rest of the paper is structured as follows. We review in section 2 the background for the compatibility formula, including the basics of topological string on toric Calabi-Yau threefolds, and how the compatibility formula is required from the conjectural quantization condition for CIS and the conjectural Fredholm determinant in the spectral theory. We also review the $Y^{N,m}$ geometry and the $K$-theoretic blowup equations. Section 3 is the core of the paper where the connection between the compatibility formula and the blowup equations is established. We give some comments on the proof in section 4. We also clarify in the appendix our convention for the Nekrasov partition function for $SU(N)$ pure SYM and the refined topological string partition function for $Y^{N,m}$ geometries.

\section{Backround}
	\label{sc:bgd}
	
	\subsection{Refined topological string theory on toric Calabi-Yau}
	\label{sc:toric}
	
	We review here very briefly the basic knowledge of toric geometry and refined topological string theory we will need in this paper. The standard references on toric Calabi-Yau are \cite{hv,Cox:2000vi,hkt, Bat}. For the convention of topological string theory we follow \cite{cgum}.
	
	A toric Calabi-Yau threefold $X_\Sigma$ is completely described by its toric fan $\Sigma$. Let the number of 1-cones $\bar{v}_\alpha$ in $\Sigma$ be $n_\Sigma +3$, the Calabi-Yau condition demands there are $n_\Sigma $ charge vectors $\ell^{(i)}$, such that
	\begin{equation}
		\sum_{\alpha=1}^{n_\Sigma +3} \ell^{(i)}_\alpha \bar{v}_\alpha = 0 \ ,\qquad i=1,\ldots, n_\Sigma  \ .
	\end{equation}
	Furthermore, one can rotate $\Sigma$ in such a way that all the 1-cones have the form
	\begin{equation}
		\bar{v}_\alpha = (1, \ell_\alpha, m_\alpha) \ ,
	\end{equation}
	i.e.\ they end on the hyperplane $(1,*,*)$. It is therefore enough to write down the planar support of the toric fan, which is a convex Newton polygon $N_\Sigma$, whose vertices are
	\begin{equation}
		v_\alpha = (\ell_\alpha, m_\alpha) \ ,\quad \alpha =1,\ldots, n_\Sigma +3 \ .
	\end{equation}
	
	The mirror curve $\cC_\Sigma$ to the Calabi-Yau can be read off from $N_\Sigma$, as the latter is precisely the Newton polygon of $\cC_\Sigma$; in other words, the equation of the mirror curve is 
	\begin{equation}
		\cC_\Sigma \; : \quad \sum_{\alpha=1}^{n_\Sigma +3} a_\alpha \re^{\ell_\alpha x + m_\alpha y} = 0 \ .
	\end{equation}
	The coefficients $a_\alpha$ of the equation modulo the $\IC^*$ actions on $\re^x, \re^y$ and the overall $\IC^*$ scaling parametrize the moduli space of $\cC_\Sigma$. Alternatively, one can use the $\IC^*$ invariant Batyrev coordinates, defined by
	\begin{equation}
		z_i = \prod_{\alpha=1}^{n_\Sigma +3} a_\alpha^{\ell_\alpha^{(i)}}  \ ,\quad i = 1,\ldots, n_\Sigma  \ .
	\end{equation}
	In addition, the coefficients $a_\alpha$ can be classified into the ``true'' moduli which correspond to integral vertices inside the Newton polygon $N_\Sigma$ and the mass parameters which correspond to integral vertices on the boundary of $N_\Sigma$\footnote{This distinction was particularly emphasized in \cite{hkrs,hkp}.}. We follow the convention of \cite{cgm2}, rename the true moduli and the mass parameters to $\kappa_j$ and $\xi_k$ respectively. The number of the true moduli is the same as the genus $g_\Sigma $ of the mirror curve, while the number of independent mass parameters, after setting three of them to 1 by the three $\IC^*$ rescalings, is denoted by $r_\Sigma $ so that $g_\Sigma +r_\Sigma  = n_\Sigma $. We notice that the definition of the Batyrev coordinates can be written as
	\begin{equation}
		-\log z_i = \sum_{j=1}^{g_\Sigma } C_{i j}\log \kappa_j + \sum_{k=1}^{r_\Sigma } \alpha_{i k} \log \xi_k \ ,
	\end{equation}
	where the $n_\Sigma \times g_\Sigma $ matrix $\bb C$ is the submatrix of minus the charge matrix $\bb L$, defined by
	\begin{equation}
		\bb L = \begin{pmatrix}
		\ell^{(1)}_1 & \ell^{(1)}_2 & \ldots & \ell^{(1)}_{n_\Sigma +3} \\
		\ell^{(2)}_1 & \ell^{(2)}_2 & \ldots & \ell^{(1)}_{n_\Sigma +3} \\
		\vdots & \vdots & \ddots &\vdots \\
		\ell^{(n_\Sigma )}_1 & \ell^{(n_\Sigma )}_2 &\ldots & \ell^{(n_\Sigma )}_{n_\Sigma +3} 
		\end{pmatrix} \ ,
	\end{equation}
	with all the columns corresponding to mass parameters removed. Since $\bb C$ is of rank $g_\Sigma $, it is  possible to perform linear transformations of rows so that the last $r_\Sigma $ rows become empty and $\bb C$ is reduced to a rank $g_\Sigma $ square matrix.
	
	The Kahler moduli $t_i$ of the Calabi-Yau threefold are related to the Batyrev coordinates of the mirror curve by the mirror map
	\begin{equation}\label{eq:mm}
		- t_i = \log z_i + \widetilde{\Pi}_i(\bb z) \ ,
	\end{equation}
	where $\widetilde{\Pi}_i(\bb z)$ is a power series with finite radius of convergence. 
	Geometrically the Kahler moduli $t_i$ are identified  with the A periods of the mirror curve. By the use of quantized curve \cite{adkmv}, one can promote the notion of period to quantum period \cite{mirmor,mirmor2,Maruyoshi:2010iu,bmt,acdkv} and the corresponding quantum mirror map is given by
	\begin{equation}\label{eq:qmm}
		- t(\hbar)_i = \log z_i +\widetilde{\Pi}_i (\bb z; \hbar) \ ,
	\end{equation}
	which reduces to the standard mirror map in the limit $\hbar\rightarrow 0$. The quantum mirror map plays an important role in both the quantization condition for CIS and the conjectural spectral theory.

	With the Calabi-Yau threefold $X_\Sigma$ as the target space, one can compute the refined topological string free energy $F_{\rm ref}(\mathbf{t};\epsilon_1, \epsilon_2)$ by the means of refined holomorphic anomaly equation \cite{hk,kw} or refined topological vertex \cite{Iqbal:2012mt,ikv}. The free energy $F_{\rm ref}(\mathbf{t};\epsilon_1, \epsilon_2)$ enjoys the following form of expansion in terms of the refinement parameters $\epsilon_1, \epsilon_2$
	\begin{equation}\label{eq:Fref-expn}
		F_{\rm ref}(\mathbf{t};\epsilon_1, \epsilon_2) = \sum_{g,n=0}^{\infty} (\epsilon_1\epsilon_2)^{g-1}(\epsilon_1+\epsilon_2)^{2n} F^{(g,n)}(\mathbf{t}) \ .
	\end{equation}
	On the other hand, in the large volume limit, $F_{\rm ref}(\mathbf{t};\epsilon_1, \epsilon_2)$ splits to the perturbative contributions and the instanton contributions.
	The perturbative contributions $F_{\rm ref}^{\rm pert}(\mathbf{t};\epsilon_1, \epsilon_2)$ has the form\footnote{The terms in the parentheses are the perturbative contributions to the prepotential, which differs from the usual form of the prepotential  for compact geometries in the linear terms in $\mathbf{t}$. The form of the prepotential presented here (also given in \cite{huang1606}) is consistent with the result from gauge theory, and it makes the formulation of the quantization condition conjecture and the spectral theory conjecture reviewed in section \ref{sc:spec} more compact.}
	\be
		F_{\rm ref}^{\rm pert}(\mathbf{t};\epsilon_1, \epsilon_2) = \frac{1}{\epsilon_1\epsilon_2}\left( \frac{1}{6} \sum_{i,j,k=1}^{n_\Sigma} a_{ijk} t_it_jt_k + 4\pi^2 \sum_{i=1}^{n_\Sigma} b_i^{\rm NS}t_i\right)  
		+ \sum_{i=1}^{n_\Sigma} b_i t_i - \frac{(\epsilon_1+\epsilon_2)^2}{\epsilon_1\epsilon_2} \sum_{i=1}^{n_\Sigma} b_i^{\rm NS} t_i \ .
	\ee
	The constants $a_{ijk}, b_i$ are classical intersection numbers and they can be computed directly from the toric data of $X_\Sigma$. The constants $b_i^{\rm NS}$ can be obtained by refined holomorphic anomaly equation \cite{hk, kw}. The instanton contributions $F_{\rm ref}^{\rm inst}(\mathbf{t},\epsilon_1,\epsilon_2)$ has the form
	\begin{equation}\label{eq:Fref-BPS}
		F_{\rm ref}^{\rm inst}(\mathbf{t},\epsilon_1,\epsilon_2) = \sum_{j_L, j_R \ge 0} \sum_{{\bf d}} \sum_{w=1}^{\infty}  (-1)^{2j_L+2j_R} N^{\mathbf{d}}_{j_L,j_R} \frac{\chi_{j_L}(q_L^w) \chi_{j_R}(q_R^w)}{w(q_1^{w/2}-q_1^{-w/2})(q_2^{w/2} - q_2^{-w/2})} e^{-w \mathbf{d}\cdot \mathbf{t}} \ .
	\end{equation}
	Here the positive integers $N^\mathbf{d}_{j_L,j_R}$ are the so-called BPS invariants, which depend on the curve class $\mathbf{d}$ in the Calabi-Yau $X_\Sigma$ and two non-negative half-integers, or spins, $j_L, j_R$. Furthermore,
	\begin{equation}
		q_{1,2} = \re^{\epsilon_{1,2}} \ , \quad q_{L,R} = \re^{\epsilon_{L,R}} \ ,\quad \epsilon_{L,R} = \frac{\epsilon_1 \mp \epsilon_2}{2} \ ,
	\end{equation}
	and $\chi_j(q)$ are characters of $SU(2)$, i.e. 
	\begin{equation}
		\chi_j(q) = \frac{q^{2j+1} - q^{-2j-1}}{q -q^{-1}} \ .
	\end{equation}
	It is an interesting fact that the BPS numbers display a checkerboard pattern; namely there exists an $n$-dimensional integer vector $\mathbf{B}$, called B-field, such that all non-vanishing BPS numbers satisfy
	\begin{equation}\label{Bdef}
		2j_L + 2j_R + 1 \equiv \mathbf{B}\cdot \mathbf{d} \mod 2 \ .
	\end{equation}
	This checkerboard pattern was first discussed in \cite{ckk} and then proved when $X_\Sigma$ is a local del Pezzo surface in \cite{hmmo}.
	For the purpose  of this paper it  is convenient to introduce   the  twisted topological string free energy   as 	\be \label{twist}
		\widehat{F}_{\rm ref}(\dt;\epsilon_1,\epsilon_2)  = F^{\rm pert}_{\rm ref}(\dt; \epsilon_1,\epsilon_2) + F^{\rm inst}_{\rm ref}(\dt + \pi \ri \dB; \epsilon_1,\epsilon_2).\ee
	
	Finally there are two important limits of the refined topological string free energy. The first is the so-called unrefined limit
	\be \label{unref}
		F_{\rm top}(\mathbf{t}; g_s) = F_{\rm ref}(\mathbf{t}; \ri g_s, -\ri g_s)  \ .
	\ee
	In the large volume limit, it splits to the perturbative contributions
	\begin{equation}
		F_{\rm top}^{\rm pert}(\mathbf{t}; g_s) = \frac{1}{6g_s^2} \sum_{i,j,k=1}^{n_\Sigma} a_{ijk}t_i t_j t_k +\sum_{i=1}^{n_\Sigma} \( b_i + \frac{4\pi^2 b_i^{\rm NS}}{g_s^2} \) t_i \ ,
	\end{equation}
	and the instanton contributions
	\begin{equation}
		F^{\rm inst}_{\rm top}(\mathbf{t}; g_s) = \sum_{j_L, j_R \ge 0} \sum_{{\bf d}} \sum_{w=1}^{\infty}  \frac{(-1)^{2j_L+2j_R}}{w}N^{\mathbf{d}}_{j_L,j_R} \frac{(2j_R+1) \sin w g_s(2j_L+1)}{(2\sin \tfrac{1}{2}w g_s)^2 \sin w g_s}e^{-w\,\mathbf{d}\cdot \mathbf{t}} \ .
	\end{equation}
	This is the case first studied in topological string theory, and it can be rigorously defined for not only non-compact but compact Calabi-Yau threefolds as well. The second limit is the Nekrasov-Shatashvili limit defined by
	\be \label{fnsrf}
		F_{\rm NS}(\mathbf{t}; \hbar) = \lim_{\epsilon_2 \rightarrow 0} \ri \epsilon_2 F_{\rm ref}(\mathbf{t}; \ri \hbar, \epsilon_2) \ ,
	\ee
	which as we reviewed in the introduction is closely related to quantum integrable systems. It also consists of the perturbative contributions
	\begin{equation}
		F^{\rm pert}_{\rm NS}(\mathbf{t};\hbar) = \frac{1}{6\hbar} \sum_{i,j,k=1}^{n_\Sigma} a_{ijk} t_it_jt_k + \(\hbar+\frac{4\pi^2}{\hbar}\)\sum_{i=1}^{n_\Sigma} b_i^{\rm NS} t_i \ ,
	\end{equation}
	and the instanton contributions
	\be   
		F_{\rm NS}^{\rm inst}({\bf t}, \hbar)= \sum_{j_L, j_R \ge 0} \sum_{{\bf d}} \sum_{w=1}^{\infty} 
		(-1)^{2j_L+2j_R} N^{\bf d}_{j_L, j_R} { \sin((2j_L+1)\hbar w/2) \sin((2j_R+1)\hbar w/2)
	\over 2 w^2\sin^3(\hbar w/2) } {\re}^{- w{\bf d}\cdot {\bf t}}. 
	\ee

	\subsection{Quantization conditions and Topological strings}
	\label{sc:spec}

	Goncharov and Kenyon \cite{Goncharov2011} gave an algorithm which constructs for each toric Calabi-Yau threefold $X_\Sigma$ based on a Newton polygon $N_\Sigma$ a quantum integrable system called the cluster integrable system (CIS). There are in the CIS $g_\Sigma$ commuting Hamiltonians and $r_\Sigma$ Casimir operators, and they can be mapped to respectively the true moduli $\kappa_j$ and the mass parameters $\xi_k$ on the mirror curve $\cC_\Sigma$. When $X_\Sigma$ engineers \cite{kkv} a 5d $\mcl N=1$ supersymmetric field theory, the integrable system underlying the field theory is precisely the CIS. Later based on previous works \cite{km,hw,ghm,gkmr}, \cite{wzh,fhma,hm} gave a conjectural exact solution to the spectral problem of any CIS using the Nekrasov-Shatashvili limit free energy of the topological string on $X_\Sigma$. The claim is that the energy levels  are solutions to the following system of $g_\Sigma$ equations
	\begin{align}  
		\sum_{i=1}^{n_\Sigma} C_{ij}\left( \frac{\partial}{\partial t_i} \widehat{F}_{\rm NS}(\widehat{\mathbf{t}}(\hbar), \hbar) + \frac{\hbar}{2\pi}\frac{\partial}{\partial t_i} \widehat{F}_{\rm NS}^{\rm inst}\left(\frac{2\pi}{\hbar} \widehat{\mathbf{t}}(\hbar), {4 \pi^2\over \hbar}\right) \right) = 2\pi \left( n_j +\frac{1}{2}\right) \ ,\nonumber\\
			n_j\in\IN_0,\; j=1,\ldots, g_\Sigma \ , \label{eq:qn-conds}
	\end{align}
	indexed by non-negative integers $n_j$ which label the excitation levels. Here the hat above 
	 $\dt(\hbar)$ means we switch on the sign
	\begin{equation}
		z_i \rightarrow (-1)^{B_i} z_i 
	\end{equation}
	in the power series part $\widetilde{\Pi}_i (\bb z; \hbar)$ of the quantum mirror map.

	At the same time, a duality between the topological string and the spectral theory (TS/ST) has been developed \cite{ghm,cgm2}.
	In this  construction  one associates  $g_{\Sigma}$ difference operators of trace class
	    \be\label{opm}\ba &  \rho_i(\tilde {\bf \kappa}_i), \quad i=1,\cdots, g_\Sigma \\
	&\tilde {\bf \kappa}_i=\{\kappa_1, \cdots, \kappa_{i-1}, \kappa_{i+1},\cdots, \kappa_{g_\Sigma}\}
	 \ea\ee
	to a toric CY manifold $X_{\Sigma}$. These operators are constructed  starting from the quantization of the mirror curve to $X_{\Sigma}$ and their spectral properties  are completely encoded in the  generalized spectral determinant defined as \cite{cgm2}
	\be \label{gsd} 
		\Xi_X (  {\boldsymbol \kappa}, \hbar)= {\rm det}\left(1+ \kappa_1 \rho_1\right) {\rm det}\left(1+ \kappa_2 \rho_2\Big|_{\kappa_1=0} \right) 
		\cdots {\rm det}\left(1+ \kappa_{g_\Sigma} \rho_{g_\Sigma} \Big|_{\kappa_1=\cdots = \kappa_{g_\Sigma-1}=0} \right). 
	\ee
	According to \cite{ghm,cgm2} this spectral determinant has an explicit expression in terms of (refined) enumerative invariants of $X_\Sigma$ and it is given by
	\be \label{sdtot}
		\Xi_X( {\boldsymbol \kappa}, \hbar)= \sum_{ {\bf n} \in \IZ^{g_\Sigma}} \exp \left( \mathsf{J}_{X}(\boldsymbol{\mu}+2 \pi \ri  {\bf n},  \hbar) \right), \quad {\bb \kappa}=\re^{\bb \mu},
	\ee
	where $\mathsf{J}_{X}$ is the topological string grand potential. The exact expression of $\mathsf{J}_{X}$ was determined in a series of works \cite{mp,hmo2,cm,hmmo,ghm,gkmr}  and it reads
	\be\label{eq:JX}
	\ba 
		\mathsf{J}_{X}({\bf t}(\hbar), \hbar)=& {t_i(\hbar) \over 2 \pi}   {\partial {F}_{\rm NS}^{\rm inst}({\bf t}(\hbar), \hbar) \over \partial t_i} 
		+{\hbar^2 \over 2 \pi} {\partial \over \partial \hbar} \left(  {{F}_{\rm NS}^{\rm inst}({\bf t}(\hbar), \hbar) \over \hbar} \right) + {\widehat{F}_{\rm top}}\left( {2 \pi \over \hbar}{\bf t}(\hbar) , {4 \pi^2 \over \hbar} \right)+A({\boldsymbol \xi}, \hbar) \ .
	\ea
	\ee
	The function $A({\boldsymbol \xi}, \hbar)$ here includes the constant map  contribution \cite{bcov}. 
	The conjecture \eqref{sdtot} has two important consequences.
	
	The first one is that \cite{ghm,cgm2} give a non--perturbative definition of  topological string  in terms of ideal Fermi gas, i.e. in terms of operator theory.
	More precisely one has
	\be \label{xitoN}
		\Xi_X( {\boldsymbol \kappa}, \hbar)=\sum_{{ N_1}\geq { 0}} \cdots \sum_{{ N_{g_\Sigma}}\geq { 0}} Z({\bf{N}}, \hbar)\kappa_1^{N_1} \cdots \kappa_{g_\Sigma}^{N_{g_\Sigma}}.
	\ee
	The expansion coefficients $Z({\bf N},\hbar)$ can be interpreted as the non--perturbative completion of topological string partition function on $X_\Sigma$ where the 't Hooft parameters play the role of the flat coordinates, and they can be identified with the partition function of an ideal Fermi gas. The  density matrix of the Fermi gas is constructed from the operators \eqref{opm} obtained by quantizing the mirror curve. By using the Cauchy identity the partition function of the  gas can be expressed as a matrix model \cite{mz,kmz} providing in this way a concrete realization of the conjecture \cite{mmopen, bkmp} (now a theorem \cite{eo-proof}).
	
	The second important consequence of \eqref{sdtot}, which is the focus of this paper, is due to the following observations. The spectrum of the operators $\rho_i$ can be computed from the vanishing loci of the generalized Fredholm determinant. And the latter vanishes when the quantum mirror curve as a difference operator has a nontrivial kernel \cite{cgm2,ghm}. On the other hand, it is observed that the quantum mirror curve to $X_\Sigma$ coincides with the quantum Baxter equation of the CIS constructed from $X_\Sigma$ \cite{fhma,huang1606,cgum}. As a consequence, the energy levels of the CIS must lie within the zero loci of the Fredholm determinant. This consistency condition leads to the following identity \cite{huang1606}
	\be\label{comp}
		\sum_{\mathbf{n}\in\mathbb{Z}^{g_\Sigma}}
		\exp\left(\ri\sum_{i=1}^{g_\Sigma}n_i \pi+\widehat{F}_{\rm top}\left(\mathbf{t}+ \ri\hbar \mathbf{C}\cdot \mathbf{n}+\frac{1}{2}\ri\hbar\,\mathbf{r},\hbar\right)
		 -\ri \sum_{i=1}^{n_\Sigma}\sum_{j=1}^{g_\Sigma}C_{ij}n_j\frac{\partial }{\partial t_i}\widehat{F}_{\text{NS}}\left(\mathbf{t},\hbar\right)\right)=0\ ,
	\ee
	from which infinitely many constraints on the refined BPS invariants of $X_\Sigma$ can be extracted. We will call \eqref{comp} the \emph{compatibility formula}. 
	The definition of the matrix $\bf C$ is given in section \ref{sc:toric} and it is completely determined by the toric data of $X_\Sigma$ as explained in \cite{kpsw}. The $g_\Sigma$-dimensional integral vector $\mathbf{r}$ is an explicit ``representation'' of the $\mathbf{B}$-field satisfying
	\begin{equation}
		\mbf r \equiv \mbf B \ , \mod (2\mbb Z)^{g_\Sigma} \ .
	\end{equation}
	We define two representations $\mbf r$ and $\mbf r'$ to be equivalent if
	\begin{equation}\label{eq:B-equiv}
		\mbf r' = \mbf r + 2\dC \cdot \dn \ ,\quad \dn \in \IZ^{g_\Sigma} \ ,
	\end{equation}	
	since it only amounts to an integral shift of the summation index $\mathbf{n}$ in \eqref{comp}.
	Note that the $\mathbf{B}$-field needs only be defined up to $(2\mbb Z)^{g_\Sigma}$ in either the conjectural quantization condition \eqref{eq:qn-conds} or the conjectural Fredholm determinant \eqref{sdtot},\eqref{eq:JX}, while the compatibility formula \eqref{comp} only holds  for some but not all concrete representations $\mbf r$ of the $\mathbf{B}$-field \cite{huang1606}. Furthermore, it was conjectured that  one can always find at least $g_\Sigma$ inequivalent representations $\mathbf{r}$ of $\dB$ with which the compatibility formula is satisfied. We call this the \emph{sufficiency  conjecture } of the compatibility formula. As a consequence, one can find enough vanishing equations for the spectral determinant \eqref{sdtot}, one for each $\mathbf{r}$, such that that  their common solutions are discrete and  coincident with the energy levels of the CIS constructed from $X_\Sigma$. 
	
	The compatibility formula together with the sufficiency conjecture implies a deep connection between the spectrum of the CIS and that of the operators constructed from the quantum mirror curve. 	
	However, 	
	the sufficiency conjecture  presented in \cite{huang1606} is constructed experimentally  case-by-case  and it was proven only for some specific  values of $\hbar$.
		 	
	 In this paper we give a precise meaning to these constraints. More precisely, when the background geometry is of type  $Y^{N,m}$, we  connect the sufficiency conjecture and the compatibility formula of \cite{huang1606} to the  $K$-theoretic blowup equation of \cite{Gottsche:2006bm,naga}. In particular this connection furnishes a complete proof of the equivalence between the two  quantization schemes, i.e.~the one  which features $S$-duality  and the one  derived from the vanishing condition of the spectral determinant.

	As an additional comment we would like to stress that the compatibility formula \eqref{comp} found in  \cite{huang1606} leads to a strong relation between  unrefined and NS free energies. From the four dimensional perspective the fact that these two limits are somehow special was also anticipated in the works of \cite{gil1,gil,kpjt,tes,llnz} where it was pointed out that, in four dimensions, they are both related to isomonodromic tau functions. 
	The fact that the blowup equations might be used as a possible explanation for the connection between the results of \cite{gil1,gil} and those of \cite{kpjt,tes,llnz}  was also  suggested  by N.~Nekrasov during J.~Teschner's talk \cite{jtalk}.  It would be interesting to see if this connection can be used to understand the operator theory interpretation behind these two limits found in  \cite{bgt, ns}.

	\subsection{$Y^{N,m}$ geometry}
	\label{sc:blowup}
	
	\begin{figure}
		\centering
		\includegraphics[width=0.65\linewidth]{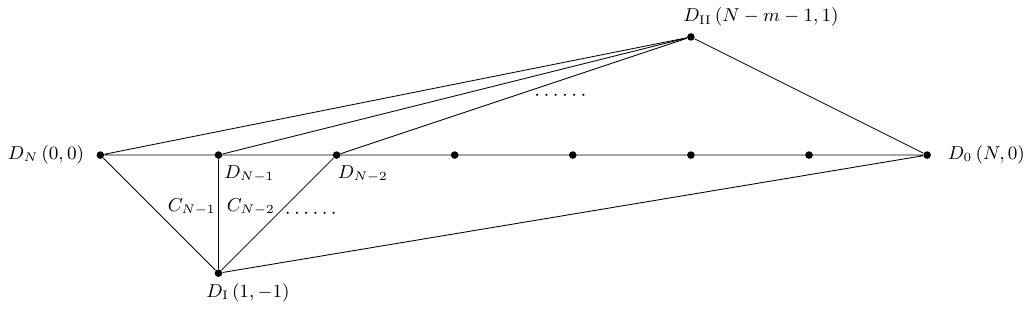}
		\caption{The planar support of the toric fan (Newton polygon) of the resolved $Y^{N,m}$ geometry.}\label{fg:Nm-fan}
	\end{figure}
	
	\begin{figure}
		\centering
		\includegraphics[width=0.6\linewidth]{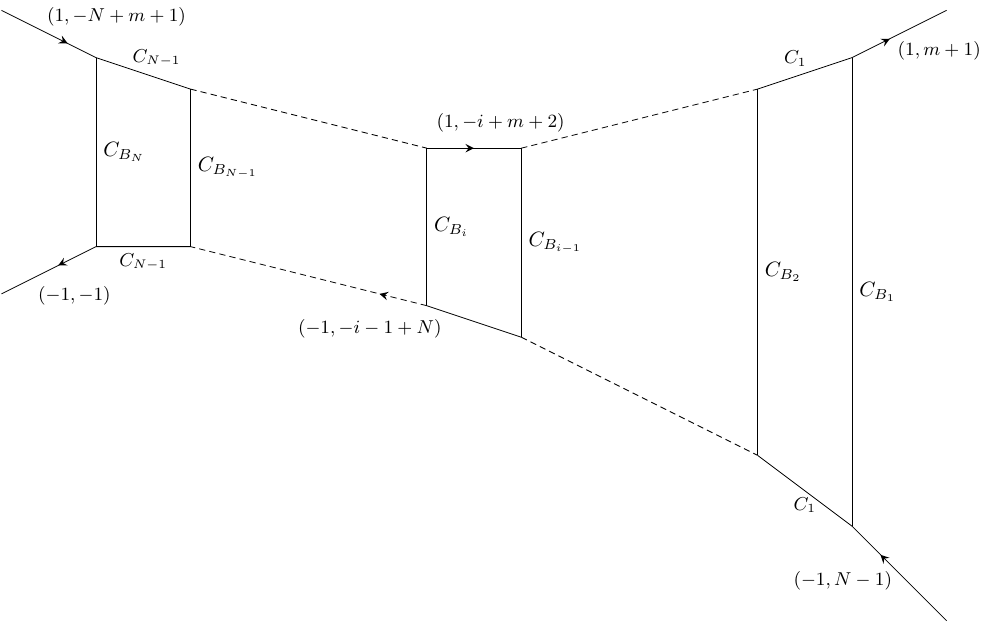}
		\caption{The dual toric diagram (ladder diagram) of the resolved $Y^{N,m}$ geometry.}\label{fg:Nm-ladder}
	\end{figure}

	The cone over the $Y^{N,m}$ singularity\footnote{This geometry was discussed in detail in \cite{Brini:2008rh}.}, or the $Y^{N,m}$ geometry for short, is a toric Calabi-Yau threefold, and it describes the fibration of the ALE singularity of $A_{N-1}$ type over $\IP^1$. Under the minimal resolution, each fiber contains an $A_{N-1}$ sphere tree. The index $m$ is an integer in the range $0\leq m \leq N$. This geometry is particularly interesting because when set up as the target space of topological string theory  it ``geometrically engineers'' the 5d $\CN=1$ pure $SU(N)$ gauge theory with Chern-Simons invariant $m$ compactified on a circle $S^1$ \cite{Iqbal:2003zz}. In particular, the refined topological string partition function can be identified with the instanton partition function of the gauge theory on the $\Omega$ background \cite{n}.
	
	\begin{figure}
		\centering
		\subfloat[$N-m$ odd\label{fg:Nmodd}]{\includegraphics[width=0.43\linewidth]{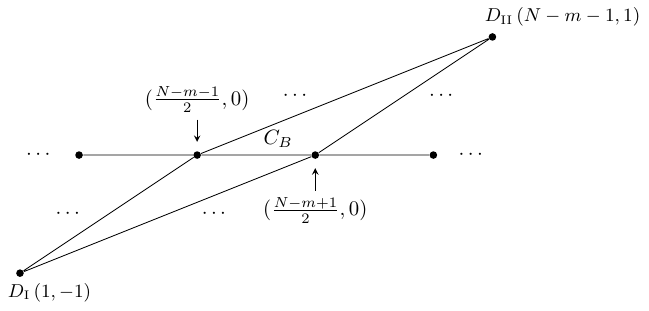}}\hspace{1ex}
		\subfloat[$N-m$ even\label{fg:Nmeven}]{\includegraphics[width=0.43\linewidth]{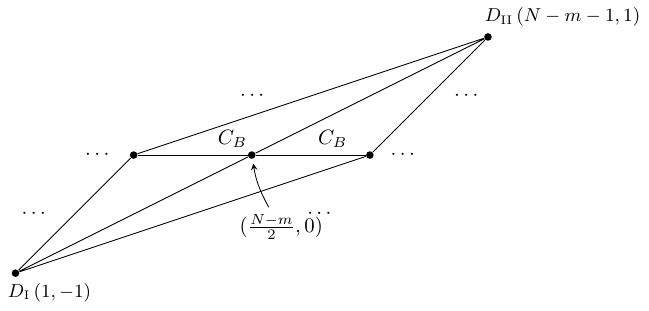}}
		\caption{The location of the edge representing the curve class $C_B$ in the Newton polygon of the resolved $Y^{N,m}$ geometry.}
	\end{figure}		
	
	The planar support of the toric fan or the Newton polygon of the resolved $Y^{N,m}$ geometry is given in Fig.~\ref{fg:Nm-fan} and the dual toric diagram (ladder diagram) in Fig.~\ref{fg:Nm-ladder}. The edges in the Newton polygon connecting the vertices $D_i$ with $i=1,\ldots, N-1$ to either the vertex $D_{\rm I}$ or $D_{\rm II}$ (these two cases are equivalent), or equivalently the edges on the two sides of the ladder diagram, correspond to the $\IP^1$'s in the $A_{N-1}$ sphere tree. We denote these curve classes by $C_i$ and the Kahler moduli that measure their sizes by $t_i$ with $i=1,\ldots, N-1$. The edges in the middle of the Newton polygon connecting neighboring vertices $D_i$  and $D_{i+1}$ with $i=0,\ldots, N-1$, or equivalently the parallel rungs in the ladder diagram, do not correspond to linearly independent curve classes. We can take the shortest rung in the ladder diagram to represent the base $\IP^1$. We denote its curve class by $C_B$ and the Kahler modulus measuring its size by $t_B$. Depending on whether $N-m$ is odd or even, the location of the edge corresponding to $C_B$ in the Newton polygon is displayed in either Fig.~\ref{fg:Nmodd} or Fig.~\ref{fg:Nmeven}. $\{t_1,\ldots,t_{N-1},t_B\}$  are all the independent Kahler moduli of the geometry. 
	
	The charge matrix (or intersection matrix) of the resolved $Y^{N,m}$ geometry is given in Tab.~\ref{tb:YNm-charges}. Each entry $\ell^{(i)}_j$ is the intersection number of the curve class $C_i$ and the divisor $D_j$. The submatrix consisting of the first $N-1$ rows and $N-1$ columns is minus the Cartan matrix of the $su(N)$ algebra. The charge vector associated to the curve class $C_B$ depends on whether $N-m$ is odd (first row in the table) or even (second row in the table).  For later convenience, we would like to use linear combinations of the $N$ curve classes to construct a new curve class which  does not intersect with the first $N-1$ divisors. We  normalize it so that its intersection number with the divisor $D_{\rm I}$ is 1. The result is the curve class $C_N$ whose intersection numbers are attached in the last row of Tab.~\ref{tb:YNm-charges}. Its Kahler modulus $t_N$ is related to the other Kahler moduli by
	\begin{equation}\label{eq:tN-tB}
		t_N = t_B -  \sum_{i=1}^{\lfloor \tfrac{N+m-1}{2}\rfloor} (i-i\,  m/N) \, t_i - \sum_{i=\lfloor \tfrac{N+m+1}{2} \rfloor}^{N-1} (N+m-i - i\, m/N) \, t_i   \ .
	\end{equation}
	$\{t_1,\ldots,t_{N-1},t_N\}$ also provides a basis of Kahler moduli. Here $t_N$ is an integral flat coordinate only when $m=0$, in which case the perturbative contributions to the refined topological string free energy can be written as \cite{Nekrasov:2003rj}
	\begin{equation}\label{fpertSN}
		F^{\rm pert}_{\rm ref}(\dt; \epsilon_1,\epsilon_2) = \frac{1}{\epsilon_1\epsilon_2}\sum_{1\leq i< j\leq N}\left(\frac{t_{ij}^3 }{6} + \frac{t_N}{2N} t_{ij}^2 \right)+ \frac{\epsilon_1^2+\epsilon_2^2 + 3\epsilon_1\epsilon_2 - 4\pi^2}{\epsilon_1\epsilon_2}\sum_{i=1}^{N-1} \frac{i(N-i)}{12}t_i  \ ,
	\end{equation}
	where
	\begin{equation}
		t_{ij} = \sum_{k=i}^{j-1} t_k \ .
	\end{equation}
	\begin{table}
		\centering
		\renewcommand*{\arraystretch}{1.2}
		\begin{tabular}{*{13}{>{$}c<{$}}}
			& D_1 & D_2 & \cdots & \cdots & \cdots &  \cdots & \cdots & D_{N-1} & D_0 & D_N & D_{\rm I} & D_{\rm II}\\\toprule
			C_1 & -2 & 1 & 0 & \cdots & \cdots & \cdots & 0 & 0 & 1 & 0 & 0 & 0 \\
			C_2 & 1 & -2 & 1 & \cdots & \cdots & \cdots & 0 & 0 & 0 & 0 & 0 & 0 \\
			C_3 & 0 & 1 & -2 & \cdots & \cdots & \cdots & 0 & 0 & 0 & 0 & 0 & 0 \\
			\vdots & \vdots & \vdots & \vdots & \vdots & \vdots & \vdots & \vdots & \vdots & \vdots & \vdots & \vdots & \vdots\\
			C_{N-2} & 0 & 0 & 0 & \cdots & \cdots & \cdots & -2 & 1 & 0 & 0 & 0 & 0 \\
			C_{N-1} & 0 & 0 & 0 & \cdots & \cdots & \cdots & 1 & -2 & 0 & 1 & 0 & 0 \\\hdashline
			\multirow{2}{*}{$C_{B}$} & \cdots & \cdots & \cdots & -1 & -1 & \cdots & \cdots & \cdots & \cdots & \cdots & 1 & 1\\
			& \cdots & \cdots & \cdots & -2 & \cdots & \cdots & \cdots & \cdots & \cdots & \cdots & 1 & 1\\\hdashline
			C_N & 0 & 0 & 0 & \cdots & \cdots & \cdots & 0 & 0 & -1-\tfrac{m}{N} & -1 + \tfrac{m}{N} & 1 & 1 \\\bottomrule
		\end{tabular}
		\caption{The charge (or intersection) matrix of the resolved $Y^{N,m}$ geometry. Each row corresponds to a curve class and each column a divisor class. When $N-m$ is odd, the intersection numbers $-1$ of $C_B$ (first row) are in the columns of  $D_{(N-m-1)/2}$ and $D_{(N-m+1)/2}$, while when $N-m$ is even , the intersection number $-2$ of $C_B$ (second row) is  in the column of $D_{(N-m)/2}$. All undisplayed numbers in these two rows are zeros. The curve class $C_N$ is not linearly independent from the others.}\label{tb:YNm-charges}
	\end{table}

\subsection{$K$-theoretic blowup equations}
	
	The blowup equations were developed by mathematicians to study the Donaldson-Witten theory \cite{budt}, and they have since borne many fruitful results such as  \cite{Gottsche:aa, Lossev:1997bz, Losev:1997tp, Edelstein:2000aj,Gottsche:1996aoa,Marino:1998bm,Takasaki:1999nv}. For instance, the celebrated pure $SU(2)$ Seiberg-Witten curve first appeared in \cite{budt}, before it was independently proposed and thoroughly studied in the seminar works of Seiberg and Witten \cite{Seiberg:1994aj,Seiberg:1994rs}. 


	In this paper we focus on the $K$-theoretic blowup equations developed in \cite{Gottsche:2006bm,Nakajima:2003pg,naga,ny,Nakajima:2011} and we give a short review here.

	It is generally difficult to compute the instanton partition function of 4d $\CN=2$ gauge theory with gauge group $SU(N)$ on $\IC^2 =\IR^4$, because the moduli space of $n$ instantons suffers from both infrared and UV divergences. The UV divergence can be cured by considering instead  the moduli space $\CM(N,n)$ of non-commutative instantons. Mathematically speaking, $\CM(N,n)$ is the
	moduli space of framed torsion-free sheaves $(F,\Phi)$ on $\IP^2$, where the framing $\Phi$ is the trivialization of the sheaf $F$ on the line at infinity. The infrared divergence on the other hand is regulated by the $\Omega$-background parametrized by $(\epsilon_1, \epsilon_2)$. To be more precise, the space $\IC^2$ has a natural action of the group $SO(4)$
	\begin{equation}
		(\epsilon_1, \epsilon_2) \; : \;\; (x,y) \in \IC^2 \mapsto (\re^{\epsilon_1} x , \re^{\epsilon_2} y) \ ,
	\end{equation}
	which induces a group action on $\CM(N,n)$. The instanton partition function is then computed by the integration of the equivariant cohomology class $\bb 1$ over $\CM(N,n)$
	\begin{equation}\label{eq:Nek-4d}
		Z^{\rm inst}(\epsilon_1, \epsilon_2,\vec{a}; \fq) = \sum_{n=0}^\infty \fq^n \int_{\CM(N,n)} \bb 1 \ .
	\end{equation}
	 We denote by $ \fq$  the fugacity of the instanton counting while
 $\vec{a}$ is a vector in the Coulomb branch. It is considered an element of the Cartan subalgebra of $su(N)$ with the constraint $\sum_{\alpha=1}^N a_\alpha = 0$; so it can be decomposed by
	\begin{equation} \label{roots}
		\vec{a} = (a_1,\ldots,a_N) =\sum_{\alpha=1}^N a_{\alpha}\vec{e}_{\alpha}= \sum_{i=1}^{N-1} a_i \,\vec{\omega}_i \ , \quad a_i =\langle \vec{a}, \vec{\alpha}^\vee_i \rangle ~,
	\end{equation}
	where $\vec{\omega}_i$ are the fundamental weights of $su(N)$ while $\vec{\alpha}^\vee_i$ denote the coroots. We will use the following conventional basis for the coroots of $su(N)$
	\begin{equation}\label{eq:coroots-basis}
		\vec{\alpha}^\vee_i = -\vec{e}_i + \vec{e}_{i+1} \ .
	\end{equation}
 Generalizing \eqref{eq:Nek-4d}, it is suggested in \cite{n} to also consider the integration in equivariant $K$-theory. The result, called the $K$-theoretic Nekrasov partition function, formally reads
	\begin{equation}
		Z^{\rm inst}(\epsilon_1, \epsilon_2,\vec{a}; \fq, \beta) = \sum_{n=0}^\infty \left(\fq \beta^{2N} e^{-N\beta(\epsilon_1+\epsilon_2)/2} \right)^n Z_n(\epsilon_1,\epsilon_2,\vec{a};\beta) \ .
	\end{equation}
	It returns to the result of equivariant cohomology when the formal $K$-theoretic parameter $\beta$ goes to 0. From the physics point of view, the $K$-theoretic Nekrasov partition function is desirable as it computes the partition function of 5d $\CN=1$ gauge theory compactified on $S^1$ on the $\Omega$-background with trivial Chern-Simons invariant $m=0$, while the latter as we have mentioned before is identified with the refined partition function of the topological string theory with the target space the $Y^{N,0}$ geometry. Since we are interested in the application in topological string theory instead of the 4d limit, we will only be concerned with the $K$-theoretic results. We can set $\beta=1$ and suppress its appearance. Hopefully without causing confusion, from now on $Z^{\rm inst}(\epsilon_1,\epsilon_2,\vec{a};\fq)$ without $\beta$ stands for the $K$-theoretic/5d partition function.
	
	When the $\Omega$-background parameters $\epsilon_1, \epsilon_2$ are turned on, the computation of $Z^{\rm inst}(\epsilon_1, \epsilon_2,\vec{a}; \fq)$ is greatly simplified as the contributing instantons are localized at the fixed point of the group action $(x,y)= (0,0)$, and they are classified by an $N$-tuple of Young diagrams $\vec{\mu} = (\mu_1, \mu_2, \ldots, \mu_N)$. 
	
	Similarly one can consider the instanton partition functions on the blowup of $\IC^2$ at the origin. The smooth moduli space $\CM(N,k,n)$ of the non-commutative instantons now is labeled not only by $N, n$, but in addition by an integer $k$ which is the magnetic flux through the exceptional divisor $E$. The natural action of the group $SO(4)$ on the blowup space $Bl_1(\IC^2)$ is now
	\begin{equation}
		(\epsilon_1, \epsilon_2) \; : \;\; (x,y), [z, w] \in Bl_1(\IC^2) \mapsto (e^{\epsilon_1} x, e^{\epsilon_2} y), [e^{\epsilon_1} z, e^{\epsilon_2}w ]
	\end{equation}
	where $z, w$ are the homogeneous coordinates on the exceptional divisor $\IP^1$. Nakajima and Yoshioka showed \cite{naga} that one can construct certain equivariant $K$ theoretic correlation function $\widehat{Z}^{\rm inst}_{k,d}(\epsilon_1,\epsilon_2,\vec{a};\fq)$ associated to the degree $d$ line bundle $\cl O(d E)$, such that it either coincides with $Z^{\rm inst}(\epsilon_1,\epsilon_2,\vec{a};\fq)$ or vanishes depending on the values of $d$ and $k$. More precisely,
	\begin{equation}\label{eq:Zhat-vanishing}
		\widehat{Z}^{\rm inst}_{k,d}(\epsilon_1,\epsilon_2,\vec{a};\fq) = 
		\begin{cases}
			Z^{\rm inst}(\epsilon_1, \epsilon_2, \vec{a}; \fq)  & 0\leq d \leq N ,\, k=0 \ ,\\
			0 & 0<k,d< N \ .
		\end{cases}
	\end{equation}
	On the other hand, using again the Atiyah-Bott-Lefschetz localization formula \cite{ab:1966}, the contributing instantons to $\widehat{Z}^{\rm inst}_{k,d}(\epsilon_1,\epsilon_2,\vec{a};\fq)$ are localized to the fixed loci consisting of two points
	\begin{equation}
		(0,0), [0,1] \quad\textrm{and}\quad (0,0),[1,0] \ ,
	\end{equation}
	and they are classified by two $N$-tuples of Young diagrams as well as an $N$-dimensional vector $\vec{k}$ satisfying $\{\vec{k}\} = -k/N$, which means
	\begin{equation} \label{kconst}
		\vec{k} \in\{ (k_\alpha) = (k_1,k_2,\ldots,k_N) \in \IQ^N \Big| \sum_{\alpha} k_\alpha = 0, \forall \alpha \, k_\alpha \equiv -k/N \mod 1 \} \ .	
	\end{equation}
	Then one can show that $\widehat{Z}^{\rm inst}_{k,d}(\epsilon_1,\epsilon_2,\vec{a};\fq)$ can be expressed in terms of two copies of the $K$-theoretic Nekrasov partition function on $\IC^2$
	\begin{equation}\label{eq:Zhat-formula}
	\begin{aligned}
		\widehat{Z}_{k,d}(\epsilon_1,\epsilon_2,\vec{a};\fq) = \exp&\left[-\frac{(4d-N)(N-1)}{48} (\epsilon_1+\epsilon_2)  \right] \\
		\times \sum_{\{\vec{k}\}=-k/N} & Z\left(\epsilon_1,\epsilon_2-\epsilon_1,\vec{a}+ \epsilon_1 \vec{k} ; \exp\left( \epsilon_1 (d - \tfrac{N}{2})\right) \fq \right) \\
		\times & Z\left(\epsilon_1-\epsilon_2,\epsilon_2,\vec{a}+ \epsilon_2 \vec{k} ; \exp\left( \epsilon_2 (d- \tfrac{N}{2})\right) \fq \right) \ .
	\end{aligned}	
	\end{equation}
	Here we have used the \emph{full} Nekrasov partition functions defined by
	\begin{equation}
	\begin{aligned}
		Z(\epsilon_1,\epsilon_2,\vec{a};\fq)  &= \exp(-\sum_{\vec{\alpha}\in\Delta} \tilde{\gamma}_{\epsilon_1,\epsilon_2}(\langle \vec{a}, \vec{\alpha}\rangle | \fq)) Z^{\rm inst}(\epsilon_1,\epsilon_2,\vec{a};\fq) \ ,\\
		\widehat{Z}_{k,d}(\epsilon_1,\epsilon_2,\vec{a};\fq) &= \exp(-\sum_{\vec{\alpha}\in\Delta} \tilde{\gamma}_{\epsilon_1,\epsilon_2}(\langle \vec{a}, \vec{\alpha}\rangle | \fq)) \widehat{Z}^{\rm inst}_{k,d}(\epsilon_1,\epsilon_2,\vec{a};\fq) \ .
	\end{aligned}
	\end{equation}
	We denote by $\Delta$ the set of roots of $su(N)$ and the function $\tilde{\gamma}_{\epsilon_1,\epsilon_2}( \langle \vec{a} , \vec{\alpha}\rangle  | \fq)$ includes the semi-classical and the one-loop contributions to the gauge theory partition function. More precisely we have 
	\begin{equation}
		\tilde{\gamma}_{\epsilon_1,\epsilon_2}(x;\fq) = \gamma^{\rm (cls)}_{\epsilon_1,\epsilon_2}(x;\fq) + \gamma^{\rm (1-loop)}_{\epsilon_1,\epsilon_2}(x) \ ,
	\end{equation}
	where
	\begin{equation}
	\begin{aligned}
		\gamma^{\rm (cls)}_{\epsilon_1,\epsilon_2}(x;\fq) =& \frac{1}{24N\epsilon_1\epsilon_2}\log(\fq)(6x^2 + 6x(\epsilon_1+\epsilon_2) + \epsilon_1^2+\epsilon_2^2 + 3\epsilon_1\epsilon_2) \\
		&+\frac{1}{12\epsilon_1\epsilon_2} \left( \pi^2(\epsilon_1+\epsilon_2) - \frac{1}{8}(2x+\epsilon_1+\epsilon_2)^3 + 2(\pi^2 x - 6\zeta(3))\right) \ ,
	\end{aligned}	
	\end{equation}
	and
	\begin{equation}
		\gamma^{\rm (1-loop)}_{\epsilon_1,\epsilon_2}(x) = \sum_{n\geqslant 1} \frac{1}{n}\frac{e^{-n x}}{(e^{ n \epsilon_1}-1)( e^{ n \epsilon_2} - 1)} \ .
	\end{equation}
	Note that the definition of $\gamma^{\rm (1-loop)}_{\epsilon_1,\epsilon_2}(x) $ is only valid for $ x >0$. The same function in the regime of $ x<0 $ is defined via analytic continuation as explained for instance in \cite{naga}. 
	The expression \eqref{eq:Zhat-formula} above together with the vanishing condition \eqref{eq:Zhat-vanishing} gives us a set of recursion equations on the $K$-theoretic Nekrasov partition function on $\IC^2$ called the $K$-theoretic \emph{blowup equations}.
	
	We can also turn on a non-vanishing Chern-Simons invariant $m$, an integer in the range of $1\leq m \leq N$, and we use an additional subscript $m$ to distinguish the partition functions from the ones with $m=0$. With the full $K$-theoretic Nekrasov partition functions defined by
	\begin{equation}\label{eq:Zm}
	\begin{aligned}
		Z_m(\epsilon_1,\epsilon_2,\vec{a}; \fq) &= \exp\left(-\sum_{\vec{\alpha}\in \Delta} \tilde{\gamma}_{\epsilon_1,\epsilon_2}( \langle \vec{a} ,\vec{\alpha} \rangle\, | \fq ) - m \sum_{\alpha=1}^{N} \frac{a_\alpha^3}{6\epsilon_1\epsilon_2}\right)  Z_m^{\rm inst}(\epsilon_1,\epsilon_2,\vec{a};\fq) \ , \\
		\widehat{Z}_{m,k,d}(\epsilon_1,\epsilon_2,\vec{a}; \fq) &= \exp\left(-\sum_{\vec{\alpha}\in \Delta} \tilde{\gamma}_{\epsilon_1,\epsilon_2}( \langle \vec{a} ,\vec{\alpha} \rangle\, | \fq ) - m \sum_{\alpha=1}^{N} \frac{a_\alpha^3}{6\epsilon_1\epsilon_2}\right)  \widehat{Z}_{m,k,d}^{\rm inst}(\epsilon_1,\epsilon_2,\vec{a};\fq) \ ,
	\end{aligned}
	\end{equation}
	the blowup formula \eqref{eq:Zhat-formula} is generalized to \cite{Gottsche:2006bm}
	\begin{equation}\label{eq:blowup-Zm}
	\begin{aligned}
		\widehat{Z}_{m,k,d}(\epsilon_1,\epsilon_2,\vec{a};\fq) =& \exp\left[\left( -\frac{(4(d+m(-\tfrac{1}{2}+\tfrac{k}{N}))-N)(N-1)}{48}  +\frac{k^3 m}{6N^2} \right) (\epsilon_1+\epsilon_2)  \right] \\
		\times \sum_{\{\vec{k}\}=-k/N} &Z_m\left(\epsilon_1,\epsilon_2-\epsilon_1,\vec{a}+ \epsilon_1 \vec{k} ; \exp\left( \epsilon_1 (d+m(-\tfrac{1}{2}+\tfrac{k}{N}) - \tfrac{N}{2})\right) \fq \right) \\
		\times &Z_m\left(\epsilon_1-\epsilon_2,\epsilon_2,\vec{a}+ \epsilon_2 \vec{k} ; \exp\left( \epsilon_2 (d+m(-\tfrac{1}{2}+\tfrac{k}{N}) - \tfrac{N}{2})\right) \fq \right) \ .
	\end{aligned}	
	\end{equation}
	The proof for the vanishing conditions \eqref{eq:Zhat-formula} can also be generalized \cite{Nakajima:2011}\footnote{We thank Mikhail Bershtein for pointing out this reference to us.}. Using the second line of \eqref{eq:Zhat-vanishing} we arrive at a subset of the $K$-theoretic blowup equations for generic values of $m$
	\begin{equation}\label{bu}
	\begin{aligned}
		0=\sum_{\{\vec{k}\}=-k/N} &Z_m\left(\epsilon_1,\epsilon_2-\epsilon_1,\vec{a}+ \epsilon_1 \vec{k} ; \exp\left( \epsilon_1(d+m(-\tfrac{1}{2}+\tfrac{k}{N}) - \tfrac{N}{2})\right) \fq \right) \\
		\times &Z_m\left(\epsilon_1-\epsilon_2,\epsilon_2,\vec{a}+ \epsilon_2 \vec{k} ; \exp\left( \epsilon_2 (d+m(-\tfrac{1}{2}+\tfrac{k}{N}) - \tfrac{N}{2})\right) \fq \right) \ ,
	\end{aligned}	
	\end{equation}
	for $0<k,d<N$.
	We will demonstrate that the sufficiency conjecture of the compatibility formula \eqref{comp} is a natural consequence of the blowup equation \eqref{bu}.

	\section{Blowup equations and quantization conditions}
	
	In this section we derive, from the $K$-theoretic blowup equations \cite{naga,Gottsche:2006bm,Nakajima:2011},  the compatibility formula \eqref{comp} between the S-invariant quantization condition of the cluster integrable systems \cite{wzh,hm,fhma} and the spectral theory associated to toric Calabi-Yau threefolds \cite{ghm,gkmr,cgm2} for the $Y^{N,m}$ geometries,  thus proving the latter.

	\subsection{Preparation: the Nekrasov partition function}
	In this section we write the Nekrasov partition function \eqref{eq:Zm} in a form which is suitable for the purpose of this paper.

	The $\mathbf{B}$-field for the $Y^{N,m}$ geometry, corresponding to the Kahler moduli $\{t_1,\ldots,t_{N-1},t_B\}$, is known to have the form \cite{hm,huang1606}
	\begin{equation}\label{eq:B-Nm}
		\dB \equiv (0, \ldots, 0, N+m) \mod (2\mbb Z)^{N-1} \ .
	\end{equation}
	Since in the twisted free energy \eqref{twist} the $\mathbf{B}$-field appears in the exponential, we can choose any representation of it, for instance the one on the right hand side of \eqref{eq:B-Nm}.
	
	We start  by writing the semiclassical and one-loop contributions from $\tilde{\gamma}_{\epsilon_1,\epsilon_2}(\langle \vec{a}, \vec{\alpha} \rangle |\fq)$ in a form which will be useful later. The vector $\vec{\alpha}$ denotes a root of $su(N)$ and, since these roots come in pairs, it is convenient to consider 
	\begin{equation}
		\tilde{\gamma}_{\epsilon_1,\epsilon_2}( x|\fq ) + \tilde{\gamma}_{\epsilon_1,\epsilon_2}( -x| \fq ) \ ,\quad x = \langle \vec{a}, \vec{\alpha} \rangle, \;\vec{\alpha} \in \Delta_+ \ .
	\end{equation}
	For the semi-classical contributions it is easy to see that 
	\be\label{dgamma2}
	\ba
		\gamma^{\rm (cls)}_{\epsilon_1, \epsilon_2}(x | \fq)+\gamma^{\rm (cls)}_{\epsilon_1, \epsilon_2}(-x| \fq)
		=&\frac{1}{\epsilon_1\epsilon_2}\left({x^2\over 2N} \log(\fq) - {2\zeta(3)}\right) +\frac{\epsilon_1+\epsilon_2}{\epsilon_1\epsilon_2}\left(\frac{\pi^2}{6} - \frac{x^2}{4}\right)\\
		& + \frac{\epsilon_1^2+\epsilon_2^2+3\epsilon_1\epsilon_2}{12 N\epsilon_1\epsilon_2}\log (\fq) - \frac{(\epsilon_1+\epsilon_2)^3}{48\epsilon_1\epsilon_2}  \ .
	\ea
	\ee
	As for $\gamma^{\rm (1-loop)}_{\epsilon_1, \epsilon_2}(x)$, it can be written as \cite{naga}
	\begin{equation}
		\gamma^{\rm (1-loop)}_{\epsilon_1,\epsilon_2}(x)   = \sum_{\ell = 0}^{\infty} C_\ell \,  \Li_{3-\ell}(e^{- x}) \ .
	\end{equation}
	where
	\begin{equation}
		C_\ell = \sum_{m=0}^{\ell} \frac{\epsilon_1^{m-1} \epsilon_2^{\ell-m-1}B_{m}B_{\ell-m}}{m!(\ell-m)!} \ .
	\end{equation}
	Therefore it can be analytically continued to the regime of $x<0$, using
	\begin{equation}
		\Li_n(z) + (-1)^n \Li_n(1/z) = -\frac{(2\pi\ri)^n}{n!} B_n\(\frac{1}{2} + \frac{\log(-z)}{2\pi\ri}\) \ ,\quad\quad z\not\in\, ]0;1] \ ,
	\end{equation}
	where $B_n(x)$ are Bernoulli polynomials. As a result, we find
	\be\label{dgamma1}
	\ba
		\gamma^{\rm (1-loop)}_{\epsilon_1,\epsilon_2}(x) + \gamma^{\rm (1-loop)}_{\epsilon_1,\epsilon_2}(-x) &=  F^{ \rm (1-loop)}  (\epsilon_1, \epsilon_2, x)\\
		&- \sum_{\ell=0}^{3} C_\ell \frac{(2\pi\ri)^{3-\ell}}{(3-\ell)!} B_{3-\ell}\left(\frac{1}{2} + \frac{\log(-e^{ x})}{2\pi\ri}\right) \ ,
	\ea
	\ee	
	where we define
	\be \label{eq:F-1loop}
		F^{\rm (1-loop)} (\epsilon_1, \epsilon_2, x) = \sum_{n=1}^\infty \frac{1}{n} \frac{e^{(\epsilon_1+\epsilon_2)n/2}+{e^{-(\epsilon_1+\epsilon_2)n/2}}}
		{(e^{\epsilon_1 n/2} -e^{-\epsilon_1 n/2})(e^{\epsilon_2 n/2} -e^{-\epsilon_2 n/2})} e^{-n x} \ .
	\ee
	In \eqref{dgamma1} we can choose
	\begin{equation}\label{eq:phase}
		\log (- \re^x) = x + \pi\ri \ ,
	\end{equation}
	and we will see that the choice of the branch cut of $\log$ here does not change the final result.
	
	Combining \eqref{dgamma2}, \eqref{dgamma1}, we find
	\begin{equation}\label{eq:Zm-split}
		\log Z_m(\epsilon_1,\epsilon_2,\vec{a}; \fq) = F_m(\epsilon_1,\epsilon_2, \vec{a};\fq) + F_{\rm aux}(\epsilon_1,\epsilon_2,\vec{a};\fq) \ ,
	\end{equation}
	where

	\begin{align}
		F_m(\epsilon_1,\epsilon_2,\vec{a};\fq) =& \;\frac{1}{\epsilon_1 \epsilon_2} \sum_{1\leq i< j \leq N}\(\frac{a_{ij}^3}{6} - \frac{\log(e^{-N \pi\ri}\fq)}{2N}a_{ij}^2 - 4\pi^2\frac{a_{ij}}{12} \) \nn
		& + 
		\frac{\epsilon_1^2 + \epsilon_2^2+3\epsilon_1\epsilon_2}{\epsilon_1\epsilon_2}\sum_{1\leq i < j \leq N}\frac{a_{ij}}{12} - \frac{m}{6\epsilon_1\epsilon_2} \sum_{\alpha=1}^{N} a^3_{\alpha} \nn
		&+\sum_{1\leq i < j \leq N} F^{\rm (1-loop)}(\epsilon_1,\epsilon_2,a_{ij})+ \log Z_m^{\rm inst}(\epsilon_1,\epsilon_2,\vec{a};\fq)  \ , \label{eq:Fm-pert}\\
		F_{\rm aux}(\epsilon_1,\epsilon_2,\vec{a};\fq) = &\sum_{1\leq i < j \leq N} 
		\( \frac{2\zeta(3)}{\epsilon_1\epsilon_2} - \frac{\epsilon_1+\epsilon_2}{\epsilon_1\epsilon_2} \pi \ri \frac{a_{ij}}{2} \right. \nn
		&\left. - \frac{\epsilon_1^2 + \epsilon_2^2 + 3\epsilon_1\epsilon_2}{24N\epsilon_1 \epsilon_2}\log(e^{-N \pi\ri }\fq) + \frac{\epsilon_1^3 + \epsilon_1^2\epsilon_2 + \epsilon_1\epsilon_2^2 + \epsilon_2^3}{48\epsilon_1\epsilon_2}\) \ .
	\end{align}
	Here we use the notation
	\begin{equation}
		a_{ij} = \sum_{k=i}^{j-1} a_k \ .
	\end{equation}
	Now we claim that by using the dictionary
	\begin{equation}\label{eq:dict}
	\left\{\begin{aligned}
		& a_i = t_i \ ,\\
		& \log \fq = -t_N + N\pi \ri \ ,
	\end{aligned}\right.
	\end{equation}
	we have the identification
	\begin{equation}\label{eq:Fm-Fref}
		F_m(\epsilon_1,\epsilon_2,\vec{a};\fq) \cong \widehat{F}_{\rm ref}(\dt;\epsilon_1,\epsilon_2) \ ,
	\end{equation}
	where $\cong$ means identification up to a term proportional to $ t_N$. This notation is convenient since this kind of terms can be factored out of the sum appearing in the compatibility formula \eqref{comp} and are therefore irrelevant.
	
	The identity \eqref{eq:Fm-Fref} can be shown as follows. From \cite{taki} it is known that one can identify the last line of \eqref{eq:Fm-pert} with the refined topological string free energy on the $Y^{N,m}$ geometry  using the dictionary \eqref{eq:dict} together with the relation between $t_N$ and $t_B$ \eqref{eq:tN-tB},
	 i.e. 
	\begin{equation}\label{eq:Zref-inst}
		\exp\(\sum_{1\leq i < j \leq N} F^{\rm (1-loop)}(\epsilon_1,\epsilon_2,a_{ij})\) Z_m^{\rm inst}(\epsilon_1,\epsilon_2,\vec{a};\fq) = Z^{\rm inst}_{\rm ref}(\dt+\pi\ri\dB;\epsilon_1,\epsilon_2) \ .
	\end{equation}
	In particular, the 1-loop corrections in the gauge theory partition function correspond to the contributions of the BPS states with spins $(j_L, j_R) = (0,1/2)$ to the refined topological string instanton partition function \eqref{eq:Fref-BPS}, and they come from M2 branes wrapping only $\mbb P^1$s in the $A_{N-1}$ sphere tree but not the base $\mbb P^1$ in the resolved $Y^{N,m}$ geometry.  The gauge theory instanton partition function instead corresponds to M2 branes wrapping at least once the base $\mbb P^1$. Note that on the right hand of the equation above, the Kahler moduli have been shifted properly by the $\mathbf{B}$-field, which we will clarify in the appendix.
	
	On the other hand, the first two lines of \eqref{eq:Fm-pert} translate to
	\begin{equation}\label{eq:Fpert-m0}
	\begin{aligned}
		F_m&(\epsilon_1,\epsilon_2,\vec{a};\fq)\big|_\textrm{first two lines} =  \\
		&\;\frac{1}{\epsilon_1 \epsilon_2} \sum_{1\leq i < j \leq N}\(\frac{t_{ij}^3}{6}+\frac{t_N}{2N}t_{ij}^2 - 4\pi^2\frac{t_{ij}}{12} \)   + 
		\frac{\epsilon_1^2 + \epsilon_2^2+3\epsilon_1\epsilon_2}{\epsilon_1\epsilon_2}\sum_{1\leq i < j \leq N}\frac{t_{ij}}{12} \\
		& - \frac{m}{6\epsilon_1\epsilon_2} \sum_{j=1}^{N} \( - \sum_{i=j}^{N-1} t_i + \frac{1}{N} \sum_{i=1}^{N-1} i\, t_i \)^3\ ,
	\end{aligned}
	\end{equation}
	where we have used that 
	\begin{equation}
		a_{\alpha} =\langle\vec{a}, \vec{e}_\alpha \rangle
		= - \sum_{i=\alpha}^{N-1}\langle\vec{a},  -\vec{e}_i + \vec{e}_{i+1} \rangle- \sum_{i=1}^{N-1} \langle\vec{a}, \vec{e}_i \rangle
		= - \sum_{i=\alpha}^{N-1}\langle\vec{a}, - \vec{e}_i + \vec{e}_{i+1} \rangle+ {1\over N}\sum_{i=1}^{N-1} i \langle\vec{a}, -\vec{e}_{i}+\vec{e}_{i+1} \rangle \ ,
	\end{equation}	
	together with $\sum_{\alpha=1}^N a_{\alpha}=0$.
		
	When $m=0$, the last line of \eqref{eq:Fpert-m0} vanishes, and \eqref{eq:Fpert-m0} is precisely the perturbative contributions to the refined topological string in \eqref{fpertSN} as expected. When $m>0$, to our knowledge there is no compact formula for $F^{\rm pert}_{\rm ref}(\dt;\epsilon_1,\epsilon_2)$ on the $Y^{N,m}$ geometry. However by comparing \eqref{eq:Fpert-m0} with the known results in the literature, for instance in Tab.~\ref{tb:Y3m} , we find perfect agreement. Notice that when $m>0$ it is more natural to use the integral flat coordinate $t_B$ instead of $t_N$. Furthermore, as explained in \eqref{eq:Fm-Fref}, when comparing the gauge theory results with the topological string results we allow the freedom of an additional term $c \,t_N$.

	\begin{table}
		\centering
		\renewcommand*{\arraystretch}{1.2}
		\resizebox{0.9\linewidth}{!}{
			\begin{tabular}{*{4}{>{$}l<{$}}}\\\toprule
				m & \textrm{from Nekrasov partition function} & \textrm{from topological string} & c \\\midrule
				\multirow{2}{*}{1} & F_0 = \frac{4}{27}\(t_1^3-t_2^3\) -\frac{1}{9}t_1 t_2(t_1 +2t_2) +\frac{1}{3}t_B(t_1^2 + t_1 t_2+ t_2^2) & \textrm{the same} & \\
				&F_1 = \(\frac{1}{6} -\frac{2}{3}c\) t_1 + \(\frac{1}{6} - \frac{4}{3}c\)t_2 + c\, t_B & F_1 = -\frac{1}{6} t_2 + \frac{1}{4}t_B & 1/4 \\\midrule
				\multirow{2}{*}{2} & F_0 = \frac{1}{27}\(8t_1^3 + t_2^3\) + \frac{1}{18}t_1 t_2(5t_1 + t_2) +\frac{1}{3}t_B(t_1^2 + t_1 t_2+ t_2^2) & \textrm{the same} & \\
				& F_1 = \(\frac{1}{6} - \frac{1}{3} c \) t_1 + \(\frac{1}{6} - \frac{2}{3}c\)t_2 + c\, t_B & F_1 = -\frac{1}{6}t_2 + \frac{1}{2}t_B & 1/2\\\bottomrule
			\end{tabular}
		}
		\caption{Comparison of the perturbative contributions to refined topological string free energies   on $Y^{3,m}$ with $m=1,2$ \cite{huang1606} and \eqref{eq:Fpert-m0}. We use the shorthand notation $F_0, F_1$ to denote $F^{(0,0), \rm pert}(\dt)$, $F^{(1,0), \rm pert}(\dt)= F^{(0,1), \rm pert}(\dt)$ extracted from \eqref{eq:Fpert-m0}.}\label{tb:Y3m}
	\end{table}		
	
	\subsection{Preparation: parametrizing the summation index vector}\label{aps}
	
	As an additional step we would like to parametrize the summation index vector $\vec{k}$ in the blowup equations \eqref{bu} in a more transparent manner.

	Recall that the summation index vector $\vec{k}$
	is subject to the constraint \eqref{kconst}, which for the sake of convenience we reproduce here
	\begin{equation}
		\vec{k} \in\{ (k_\alpha) = (k_1,k_2,\ldots,k_N) \in \IQ^N \Big| \sum_{\alpha} k_\alpha = 0, \forall \alpha \, k_\alpha \equiv -k/N \mod 1 \} \ .	
	\end{equation}	 
	As a consequence, $\vec{k}$ is an element in the weight lattice of $su(N)$. It is useful to write $\vec{k}$ as \cite{ny}
	\begin{equation}
		\vec{k} = (k_\alpha) = \(\ell_\alpha - \frac{k}{N}\) \ ,
	\end{equation}
	where
	\begin{equation}
		\vec{\ell} = (\ell_1,\ldots, \ell_N) =(\ell_\alpha) \in  \left\{  (\ell_\alpha) \in \IZ^{N}  \Big| \sum_{\alpha=1}^N \ell _\alpha = k \right\} \ .
	\end{equation}
	Using the conventional basis \eqref{eq:coroots-basis} for the coroots of $su(N)$,
	we find the following parametrization of the summation index vector $\vec{k}$
	\begin{equation}\label{eq:k-prm}
		\vec{k} = \sum_{i=1}^{N-1} k_i \, \vec{\alpha}_i^\vee = \sum_{i=1}^{N-1} \tilde{k}_i \, \vec{\omega}_i \ ,
	\end{equation}
	where we sum over the coroots $\vec{\alpha}_i^\vee$ of $su(N)$, or the fundamental weights $\vec{\omega}_i$ of $su(N)$, and we have
	\begin{equation}\label{eq:k-n}
		k_i = n_i + \frac{i k}{N} \ ,\quad n_i \in \IZ \ ,
	\end{equation}
	as well as
	\begin{equation}\label{eq:tk-n}
		\tilde{k}_i = \sum_{j=1}^{N-1} C_{ij}n_j + k \delta_{i,N-1} \ .
	\end{equation}
	Here $C_{ij}$ is the Cartan matrix of $su(N)$. Therefore, we can replace the index $\vec{k}$ by the more manageable\footnote{This is similar to the manipulation in \cite{Bonelli:2012ny} where one converts fractional cohomology classes to integral homology classes.} $\vec{n} \in \IZ^{N-1}$ and we write the blowup equation \eqref{bu} as
	\be\label{bui} 
	\ba
		\sum_{{\bf{n}} \in \IZ^{N-1}} &Z_m\left(\epsilon_1,\epsilon_2-\epsilon_1,\vec{a}+ \epsilon_1 \vec{k}({\bf{n}}) ; \exp\left( { \epsilon_1}(d+m(-\tfrac{1}{2}+\tfrac{k}{N}) - \tfrac{N}{2})\right) \fq \right) \\
		\times &Z_m\left(\epsilon_1-\epsilon_2,\epsilon_2,\vec{a}+ \epsilon_2 \vec{k}({\bf{n}}) ; \exp\left( { \epsilon_2}(d+m(-\tfrac{1}{2}+\tfrac{k}{N}) - \tfrac{N}{2})\right) \fq \right)=0 .
	\ea 
	\ee
	As we will see later, many terms in \eqref{bui} are $\bb n$- independent and can be factored out from the summation. Hence it is convenient to  introduce the symbol $\sim$ defined by 
	\be 
	F({\bf n})\sim G({\bf n}) \quad {\text {if }} \quad {F({\bf n}) \over G({\bf n}) } \quad {\text{is independent of $\bf n$} }.  
	\ee	

	\subsection{From blowup equation to compatibility formula}
	
	We demonstrate here that the compatibility formula \eqref{comp} can be derived from the blowup equation \eqref{bui} in the Nekrasov-Shatashvili limit
	\begin{equation}
		\epsilon_1 = \ri \hbar \ ,\quad \epsilon_2 \rightarrow 0 \ .
	\end{equation}
	We will use predominantly the notation $\sim$ and ignore terms which are independent of the summation index $\bf{n}$ or $\vec{k}$, and which therefore can be factored out of the summation in the blowup equation.
	
	We replace the Nekrasov partition functions in the blowup equation \eqref{bui} by the twisted topological string free energy $\widehat{F}_{\rm ref}(\dt;\epsilon_1,\epsilon_2)$ as well as the auxiliary function $F_{\rm aux}(\epsilon_1,\epsilon_2,\vec{a};\fq)$ using identifications \eqref{eq:Zm-split}, \eqref{eq:Fm-Fref}, and check the behavior of each term in the NS limit. Due to the dictionary \eqref{eq:dict}, we know that the extra term proportional to $t_N$ that may arise in the identification \eqref{eq:Fm-Fref} does not depend on the summation index and can be factored out.
		
	Furthermore most of the terms in $F_{\rm aux}(\epsilon_1,\epsilon_2,\vec{a};\fq)$ can be factored out except for
	\begin{equation}
	\sum_{\vec{\alpha} \in \Delta_+} -\frac{\epsilon_1+\epsilon_2}{\epsilon_1 \epsilon_2} \pi \ri \frac{\langle \vec{a}, \vec{\alpha} \rangle}{2} = -\frac{\epsilon_1+\epsilon_2}{\epsilon_1 \epsilon_2} \pi \langle \vec{a}, \vec{\rho}\rangle  \ ,
	\end{equation}
	where $\vec{\rho}$ is the Weyl vector. Therefore the contributions of the auxiliary function in the blowup formula in the NS limit is
	\begin{equation}\label{eq:compt-1}
	\begin{aligned}
	 & \exp\left[F_{\rm aux}(\epsilon_1,\epsilon_2-\epsilon_1,\vec{a} + \epsilon_1 \vec{k};\fq)+F_{\rm aux}(\epsilon_1-\epsilon_2,\epsilon_2,\vec{a} + \epsilon_2 \vec{k};\fq)\right]_{\textrm{NS limit}} \\
	\sim & \exp (-{\pi\ri \over 2}\sum_{1\leq i <j \leq N} \sum_{\ell =i}^{j-1}\tilde k_{\ell}) 
=	 \exp( -\pi\ri \sum_{i=1}^{N-1} n_i) \ .
	\end{aligned}
	\end{equation}
	
	Next, let us look at the contributions of $\widehat{F}_{\rm ref}(\mathbf{t},\epsilon_1,\epsilon_2)$. In the  first term of the blowup equation \eqref{bui} it becomes
	\begin{equation}\label{eq:compt-2}
	\begin{aligned}
		&\exp\left[\widehat{F}_{\rm ref}\(t_i+\epsilon_1 \tilde{k}_i,t_N - \epsilon_1(d+m(-\tfrac{1}{2}+\tfrac{k}{N})-\tfrac{N}{2}) ; \epsilon_1,\epsilon_2 - \epsilon_1\)\right]_{\textrm{NS limit}} \\
		\sim& \exp\left[\widehat{F}_{\rm top}\(t_i + \ri\hbar(\sum_{j=1}^{N-1} C_{ij}n_j +2r_i), t_N + 2\ri\hbar\, r_N , \hbar \) \right] \ .
	\end{aligned}
	\end{equation}
	Here the subscript $i$ of $t_i$ runs from $1$ to $N-1$. The vector $\mbf r = (r_1,\ldots,r_{N-1},r_N)$ is
	\begin{equation}\label{eq:B-N}
	r_i = \begin{cases}
		0 \ , & i\leq N-2 \\
		2k \ ,& i=N-1 \\
		N-2d- 2m\(-\tfrac{1}{2} + \tfrac{k}{N}\)  \ , & i=N \; (t_N)\ .
	\end{cases}
	\end{equation}
	Finally in the second term of the blowup equation \eqref{bui} $\widehat{F}_{\rm ref}(\mathbf{t},\epsilon_1,\epsilon_2)$ becomes 

\be \label{eq:compt-3}
\ba 
&\exp\left[\widehat{F}_{\rm ref}\(t_i + \epsilon_2 \tilde{k}_i, t_N - \epsilon_2 (d+m(-\tfrac{1}{2}+\tfrac{k}{N})-\tfrac{N}{2});\epsilon_1 - \epsilon_2,\epsilon_2\) \right]_\textrm{NS limit}\\
\sim & \exp\left[-\ri \sum_{i,j=1}^{N-1} C_{ij} n_j \frac{\partial}{\partial t_i}\widehat{F}_{\rm NS}(\mathbf{t},\hbar) \right].
\ea
\ee
	Combining all the components \eqref{eq:compt-1}, \eqref{eq:compt-2}, and \eqref{eq:compt-3} of the blowup equation in the NS limit, we find
	\begin{align}
		\sum_{n_j\in \mbb Z} &\exp\left[  -\pi\ri \sum_{i=1}^{N-1} n_i 
		+\widehat{F}_{\rm top}\(t_i + \ri\hbar(\sum_{j=1}^{N-1} C_{ij}n_j +2r_i), t_N + 2\ri\hbar\, r_N , \hbar \)
		-\ri \sum_{i,j=1}^{N-1} C_{ij} n_j \frac{\partial}{\partial t_i}\widehat{F}_{\rm NS}(\mathbf{t},\hbar)
		\right] \nn
		&=0 \ .\label{eq:comp-res}
	\end{align}	
	Notice  that a different choice of the branch cut of $\log$ in \eqref{eq:phase} only changes the first term in the exponential in \eqref{eq:comp-res} by an integral multiple of $2\pi\ri$, and is therefore irrelevant.
	
	Moreover
	in the basis of the Kahler moduli $\{t_1,\ldots,t_{N-1},t_N\}$, the Cartan matrix $C_{ij}$ of $SU(N)$ is precisely the $\mbf C$ matrix in the compatibility formula according to Tab.~\ref{tb:YNm-charges}.  	The $\mathbf{r}$-vector given in \eqref{eq:B-N} seems to be problematic as obviously it is not integral when $m>0$. This is because we are not in the natural flat coordinates for generic values of $m$. When converted to the integral basis $\{t_1,\ldots,t_{N-1},t_B\}$ using the relation \eqref{eq:tN-tB}, the $\mathbf{r}$-vector becomes
	\begin{equation}\label{eq:B-B}
		r_i = \begin{cases}
		0 \ , & i\leq N-2 \\
		2k \ , & i=N-1 \\
		N + m - 2d - 2km + 2k\cdot \min (m+1,N-1) \ , & i=B \;(t_B) \ ,
		\end{cases}
	\end{equation}
	which is both integral and consistent with the $\mathbf{B}$-field given in \eqref{eq:B-Nm} for the $Y^{N,m}$ geometry. 

	This completes our derivation of the compatibility formula \eqref{comp} from the blowup equation \eqref{bu}. In addition, since $0<k,d<N$, equation \eqref{eq:B-B} gives us $(N-1)^2$ inequivalent sets of $\mathbf{r}$-vectors\footnote{They correspond to $(N-1)^2$ different vanishing equations of the generalized Fredholm determinant, whose solution spaces do not intersect transversely.} which over-satisfy the claim of the sufficiency conjecture.
	
\subsection{Some examples for the $\mathbf{r}$-vectors}
	
	In the following we give explicit values of the inequivalent $\mathbf{r}$-vectors \eqref{eq:B-B} in some examples and we compare them with the known results in the literature \cite{huang1606}. 
	
		{In the cases of $SU(N)$ with $m=0$, it is more convenient to use the Kahler modulus $t_N$ instead of $t_B$, and we find from \eqref{eq:B-N}
	\begin{align}
		SU(3):\quad\quad & \(\begin{array}{c} r_{N-1}\\r_N\end{array}\) = \begin{pmatrix}
		2 & 4 \\-1 & 1
		\end{pmatrix} \ ,\\
		SU(4):\quad\quad & \(\begin{array}{c} r_{N-1}\\r_N\end{array}\) = \begin{pmatrix}
		2 & 4 & 6 \\-2 & 0 & 2
		\end{pmatrix} \ ,\\
		SU(5):\quad\quad & \(\begin{array}{c} r_{N-1}\\r_N\end{array}\) = \begin{pmatrix}
		2 & 4 & 6 & 8\\-3 & -1 & 1 & 3
		\end{pmatrix}	\ .
	\end{align}
	The notation here means the $\mathbf{r}$-vectors are $\mathbf{r}=(0,\ldots,0, r_{N-1}, r_N )$, where $r_{N-1}$ can take any value in the first row, and $r_{N}$ any value in the second row. They are equivalent in the sense of \eqref{eq:B-equiv} to the $\mathbf{r}$-vectors given in \cite{huang1606}.}
	
	In the cases of $SU(3)$ with $m>0$, it is more convenient to use the Kahler modulus $t_B$ instead of $t_N$ \footnote{In these coordinates one has to consider ${\bf t}+\ri \hbar {\bf n \cdot C}$ where $\bf C$ is an $N\times (N-1)$ matrix as follows from Tab.~\ref{tb:YNm-charges}.  }. For $m=1$, the inequivalent $\mathbf{r}$-vectors are \cite{huang1606}
	\begin{equation}
		(-2,2,0), (-2,0,0), (0,-2,-2), (2,-2,-2) \ .
	\end{equation}
	With the matrix $\mbf C$ \cite{huang1606}\footnote{The $\mbf C$ matrix is transposed compared to \cite{huang1606} to be consistent with our convention. }
	\begin{equation}
		C = \begin{pmatrix}
			-2 & 1\\
			1 & -2\\
			0 & -2
		\end{pmatrix} \ ,
	\end{equation}
	these $\mathbf{r}$-vectors are equivalent to
	\begin{equation}
		(0,4,4), (0,2,4), (0,4,6), (0,2,2) \ ,
	\end{equation}
	the prediction of eq.~\eqref{eq:B-B}. For $m=2$, the inequivalent $\mathbf{r}$-vectors are \cite{huang1606}
	\begin{equation}
		(-2,2,-1), (-2,0,1), (0,-2,-1), (2,-2,-1) \ .
	\end{equation}
	With the matrix $\bf C$ \cite{huang1606}
	\begin{equation}
		C = \begin{pmatrix}
			-2 & 1\\
			1 & -2\\
			0 & -1
		\end{pmatrix} \ ,
	\end{equation}	
	they are again equivalent to the prediction of eq.~\eqref{eq:B-B}
	\begin{equation}
		(0,4,1), (0,2,3), (0,4,3), (0,2,1) \ .
	\end{equation}

	\section{Discussions}
	
	In this paper we have shown the equivalence between the NS limit of the $K$-theoretic blowup equations and the compatibility formula with the corresponding sufficiency conjecture proposed in \cite{huang1606}. In the case of the toric $Y^{N,m}$ geometries, since the $K$-theoretic blowup equations have been explicitly worked out and proved \cite{naga,Gottsche:2006bm,Nakajima:2011}, our derivation provides a complete proof of the equivalence between the quantization scheme which features $S$-duality \cite{wzh,hm,fhma} and the one derived from the vanishing condition of the spectral determinant \cite{ghm,gkmr,cgm2,cgum}.
	 From that perspective, we can interpret the NS limit of the vanishing $K$-theoretic blowup equations as a formula which expresses the non-perturbative effects in  cluster integrable system in terms of the perturbative ones. It would be interesting to understand what is, and if there is, a similar interpretation for the  NS limit of the non-vanishing $K$-theoretic blowup equations.
	
	The compatibility formula and the sufficiency conjecture were formulated in \cite{huang1606} for generic toric Calabi-Yau threefolds. To extend our derivation, explicit expressions of the $K$-theoretic blowup equations for this broad category of geometries would be needed. In principle the blowup equations can be extended to any geometry that engineers a quiver gauge theory, and from there to any toric Calabi-Yau threefold by proper blowdowns and flops \cite{Eynard:2010dh,Iqbal:2012mt}. For instance, the simplest toric Calabi-Yau, local $\IP^2$, is a blowdown of $Y^{2,1}$ (local $\IF_1$), and the $\IC^3/\IZ_5$ resolution a blowdown of $Y^{3,2}$. Once the $K$-theoretic blowup equations are known, our derivation can be easily carried over, and a compact formula for all the inequivalent $\mathbf{r}$-vectors can be obtained. It would be interesting to see if the sufficiency conjecture is always over-satisfied.

        Furthermore, the compatibility formula was presented in
        \cite{wzh, huang1606} as a generating function of constraints
        for BPS invariants, although they are not sufficient to fix
        all the BPS numbers.  On the other hand, the blowup equations,
        which can be regarded as a superset of the compatibility
        formula, were found in \cite{ks} to be powerful enough to fix
        the Nekrasov partition function recursively with respect to
        increasing instanton number (base degree in the topological
        string picture). If indeed the $K$-theoretic blowup equations
        can be extended to generic topological string, they might
        provide yet another way to compute refined BPS numbers
        \footnote{ In subsequent works
          \cite{Gu:2017ccq,Huang:2017mis,Gu:2018gmy}, by using the set
          up developed here, it has been possible to extend the
          $K$-theoretic blowup equations to other toric geometries and
          even non-toric geometries as well.}.

	\section*{Acknowledgements} 
	
	We would like to thank  Mikhail Bershtein, Giulio Bonelli, Amir-Kian Kashani-Poor,  Albrecht Klemm, Marcos Mari\~{n}o and Alessandro Tanzini for valuable discussions. We are particularly grateful to Marcos Mari\~{n}o for his comment on the possible connection between the blowup equations and the compatibility formula \cite{marinostring16}\footnote{According to Marcos Mari\~{n}o, he himself was tipped off by an anonymous referee of one of his previous papers,  probably in connection with Nikita Nekrasov's observation made at 50:30 of Joerg Teschner's talk \cite{jtalk}.}, and to him as well as Amir-Kian Kashani-Poor and Alessandro Tanzini for their careful reading of the draft. 
	We would also like to thank the organizers of the String-Math 2016 at Paris where the project was conceived. JG is supported by the grant ANR-13-BS05-0001.

	\appendix
	\section{Convention for refined topological string and Nekrasov partition functions} \label{npf}

	We give here the convention we use for the refined topological string partition function on the $Y^{N,m}$ geometry and the Nekrasov instanton partition function for the $SU(N)$ SYM with Chern-Simons invariant $m$. This enables us to explain the appearance of the twist $\pi\ri\dB$ when we identify the two partition functions in \eqref{eq:Zref-inst} using the dictionary \eqref{eq:Fm-Fref}.
	
	We first introduce some notation for the refined topological string free energy. Let
	\begin{equation}
		t = \re^{\epsilon_1} \ ,\quad  q = \re^{-\epsilon_2} \ .
	\end{equation}
	Let $\vec{\mu} = (\mu_1,\mu_2,\ldots,\mu_N)$ be an $N$-tuple of Young diagrams associated to the $N$ parallel edges of the ladder toric diagram of the $Y^{N,m}$ geometry given in Fig.~\ref{fg:Nm-ladder}. Let us introduce several functions of Young diagrams: the arm length and the leg arm in a Young diagram
	\begin{equation}
		a_\mu(\alpha,\beta) = \mu_\alpha - \beta \ , \quad l_\mu(\alpha,\beta) = \mu^t_\beta - \alpha \ ;
	\end{equation}
	besides,
	\begin{gather}
		|\mu| = \sum_{\alpha} \mu_\alpha, \quad  ||\mu||^2 = \sum_{\alpha} \mu_\alpha^2 \ , \\
		\kappa_\mu = 2\sum_{(\alpha,\beta)\in\mu}(\beta-\alpha) = |\mu| + \sum_{\alpha} (\mu_\alpha^2 - 2\alpha \mu_\alpha) \ .
	\end{gather}
	We also use the notation for an $N$-tuple of Young diagrams
	\begin{equation}
		|\vec{\mu}| = \sum_{i=1}^N |\mu_i|
	\end{equation}
	Finally we introduce the function
	\begin{equation}
		\tilde{Z}_{\mu}(t,q) = \prod_{s\in\mu}\( 1 - t^{a_{\mu}(s)+1} q^{l_{\mu}(s)}\)^{-1} \ ,
	\end{equation}
	the framing factor \cite{taki}
	\begin{equation}
		f_\mu(t,q) = (-1)^{|\mu|} t^{\tfrac{||\mu^2||^2}{2}} q^{-\tfrac{||\mu||^2}{2}} \ ,
	\end{equation}
	and the framing $n_i$ associated to each parallel edge in the ladder diagram \cite{taki}
	\begin{equation}
		n_i = -(N+m-2i+1) \ .
	\end{equation}
	Then the refined topological string partition function for $Y^{N,m}$ with non-vanishing base degree computed by the refined topological vertex \cite{Iqbal:2012mt,ikv}  is \cite{taki}
	\begin{align}
		Z_{\rm ref}^{\textrm{inst},'}&(\dt,\epsilon_1,\epsilon_2) = \sum_{\vec{\mu}} 
		\(\prod_{i=1}^N (-Q_{B_i})^{|\mu_i|} f_{\mu_i}(t,q)^{n_i} q^{\tfrac{||\mu_i||^2}{2}} t^{\tfrac{||\mu^t_i||^2}{2}} \tilde{Z}_{\mu_i} (t,q) \tilde{Z}_{\mu^t_i}(q,t)\) \nn
		&\times \prod_{1\leq i < j \leq N}\prod_{\alpha,\beta=1}^\infty 
		\frac{1-Q_{F_{ij}}t^\beta q^{\alpha-1}}{1-Q_{F_{ij}} t^{-{\mu^t_{i,\alpha}+\beta}} q^{-\mu_{j,\beta}+\alpha-1}}
		\frac{1-Q_{F_{ij}} t^{\beta-1}q^\alpha}{1-Q_{F_{ij}}t^{-\mu^t_{i,\alpha}+\beta-1} q^{-\mu_{j,\beta}+\alpha}} \ ,\label{eq:Zm-vertex}
	\end{align}
	where $Q_{B_i}$ defined by
	\begin{equation}
		Q_{B_i} =\re^{-t_{B_i}}
	\end{equation}
	measure the sizes of the curves associated to the parallel edges in the ladder diagram in the resolved $Y^{N,m}$ geometry, while $Q_{F_{ij}}$ are defined by
	\begin{equation}
		Q_{F_{ij}} = \re^{-t_{ij}} \ .
	\end{equation}
	\eqref{eq:Zm-vertex} is actually (3.5) of \cite{taki} \emph{modified} with an additional minus sign in front of $Q_{B_i}$, such that in the large fiber limit
	\begin{equation}
		0< Q_{F_i} = \re^{-t_i} \ll 1 \ ,
	\end{equation}
	each summand of \eqref{eq:Zm-vertex} has sign $(-1)^{(N+m)|\vec{\mu}|}$ in accordance with \cite{Iqbal:2003zz,hm}.
	
	For the $SU(N)$ pure SYM with Chern-Simons invariant $m$, we introduce the notation
	\begin{equation}
		Q_{ij} = \re^{-a_{ij}} = \prod_{k=i}^{j-1}\re^{- \langle \vec{a}, \vec{\alpha}^\vee_k \rangle} \ ,\quad  e_i = \re^{\langle \vec{a}, \vec{e}_i \rangle }
	\end{equation}
	such that
	\begin{equation}
		Q_{ij} = e_i e_j^{-1} \ .
	\end{equation}
	The $K$-theoretic Nekrasov instanton partition function is \cite{Gottsche:2006bm,taki}\footnote{Like in the main body of the paper we have set $\beta=1$.}
	\begin{equation}\label{eq:Zm-Nek}
		Z_m^{\rm inst}(\epsilon_1,\epsilon_2,\vec{a};\fq) = \sum_{\vec{\mu}}\frac{\fq^{|\vec{\mu}|}}{\prod_{i,j=1}^N n^{\vec{\mu}}_{ij}(t,q,Q_{ij})} \(\frac{q}{t}\)^{\tfrac{N}{2}|\vec{\mu}|} \prod_{i=1}^N e_i^{m|\mu_i|} t^{-m\tfrac{||\mu^t_i||^2}{2}} q^{m\tfrac{||\mu_i||^2}{2}} \ ,
	\end{equation}
	where, with $\mu_i = \mu, \mu_j = \nu, Q_{ij} = Q$, the weight factor $n^{\vec{\mu}}_{ij}(t,q,Q_{ij})$ is defined by\footnote{Most of the terms in the weight factor cancel, leaving only a finite number of terms remaining in the denominator.}
	\begin{equation}
		\frac{1}{n^{\vec{\mu}}_{ij}(t,q,Q)} = \prod_{\alpha,\beta =1}^{\infty} \frac{1- Q t^{\beta-1}q^\alpha}{1-Q t^{-\mu^t_\beta+\alpha-1} q^{-\nu_\alpha+\beta}} \ .
	\end{equation}
	It was shown that \eqref{eq:Zm-Nek} can be cast as \cite{taki}
	\begin{align}
		& Z_m^{\rm inst}(\epsilon_1,\epsilon_2,\vec{a};\fq) \nn
		= & \sum_{\vec{\mu}} (-1)^{N|\vec{\mu}|} \fq^{|\vec{\mu}|} 
		\(\prod_{1\leq i < j \leq N} Q_{ij}^{|\mu_i|+|\mu_j|} \)
		\(\prod_{i=1}^N e_i^{m|\mu_i|}\(\frac{q}{t}\)^{-n_i \tfrac{||\mu_i^t||^2}{2}} q^{-n_i \tfrac{\kappa_{\mu_i}}{2}} t^{\tfrac{||\mu^t_i||^2}{2}} q^{\tfrac{||\mu_i||^2}{2}} \tilde{Z}_{\mu_i}(t,q) \tilde{Z}_{\mu^t_i}(q,t) \) \nn
		&\times \prod_{1\leq i < j \leq N}\prod_{\alpha,\beta=1}^\infty \frac{1-Q_{ij}t^{\alpha-1} q^\beta}{1-Q_{ij} t^{-\mu^t_{i,\beta}+\alpha-1}q^{-\mu_{j,\alpha}+\beta}}\frac{1-Q_{ij}t^{\alpha} q^{\beta-1}}{1-Q_{ij} t^{-\mu^t_{i,\beta}+\alpha} q^{-\mu_{j,\alpha}+\beta-1} } \ .
	\end{align}
	It is easy to see that the summand is similar to that of \eqref{eq:Zm-vertex}, but it has sign $(-1)^{N|\vec{\mu}|}$ in the large fiber limit.
	
	Following the derivation of \cite{taki} one concludes that the two partition functions are the same if we set $\fq  = (-1)^m \re^{-t_N}$ and $Q_{ij} = Q_{F_{ij}}$. Using the slightly different dictionary \eqref{eq:dict}, we would have
	\begin{equation}
		Z^{\textrm{inst},'}_{\rm ref}(t_1,\ldots,t_{N-1}, t_N+(N+m)\pi\ri,\epsilon_1,\epsilon_2) = Z^{\rm inst}_{m}(\epsilon_1,\epsilon_2,\vec{a};\fq) \ ,
	\end{equation}
	which implies \eqref{eq:Fm-Fref}.
	
\bibliographystyle{amsmod}
\bibliography{biblio}

\end{document}